\documentclass[12pt]{article}
\usepackage{graphicx}
\pdfoutput=1
\usepackage{mathrsfs}
\usepackage{amsfonts}
\usepackage{color}
\usepackage{amsmath}
\usepackage{setspace}
\usepackage{multirow}
\usepackage{ntheorem}
\usepackage{pstricks}
\usepackage{fancybox}
\usepackage{multirow}
\usepackage{indentfirst}

\newcommand\pubdate{\today}

\def\support{\footnote{Corresponding author: liux3771@umn.edu}}

\def\Title#1{\begin{center} {\Large #1 } \end{center}}
\def\Author#1{\begin{center}{  #1} \end{center}}
\def\Address#1{\begin{center}{ \it #1} \end{center}}

\newenvironment{Abstract}{\begin{quotation}  }{\end{quotation}}

\def\Acknowledgements{\bigskip  \bigskip \begin{center} \begin{large}
             \bf ACKNOWLEDGEMENTS \end{large}\end{center}}

\def\beq{\begin{equation}}
\def\eeq#1{\label{#1}\end{equation}}
\def\eeqn{\end{equation}}

\def\beqa{\begin{eqnarray}}
\def\eeqa#1{\label{#1}\end{eqnarray}}
\def\eeqan{\end{eqnarray}}

\let\bar=\overbar

\def\Dslash{\not{\hbox{\kern-4pt $D$}}}
\def\dslash{\not{\hbox{\kern-2pt $\del$}}}

\def\msb{{\bar{\ssstyle M \kern -1pt S}}}

\begin{document}
\begin{titlepage}
\pubdate
\vfill
\Title{Assessing the Treatment Effect Heterogeneity with a Latent Variable}
\vfill
\Author{Yunjian Yin$^{a,b}$, Lan Liu$^{b,}$\support, and Zhi Geng$^{a}$\\}
\Address{
$^{a}${\small School of Mathematical Sciences, Peking University, Beijing 100871, China}\\
$^{b}${\small School of Statistics, University of Minnesota, Minneapolis, Minnesota 55455, USA}}
\vfill
\begin{Abstract}
The average treatment effect (ATE) is popularly used to assess the treatment effect. However, the ATE implicitly assumes a homogenous treatment effect even amongst individuals with different characteristics. In this paper, we mainly focus on assessing the treatment effect heterogeneity, which has important implications in designing the optimal individual treatment regimens and in policy making. The treatment benefit rate (TBR) and treatment harm rate (THR) have been defined to characterize the magnitude of heterogeneity for binary outcomes. When the outcomes are continuous, we extend the definitions of the TBR and THR to compare the difference between potential outcomes with a pre-specified level $c$. Unlike the ATE, these rates involve the joint distribution of the potential outcomes and can not be identified without further assumptions even in randomized clinical trials. In this article, we assume the potential outcomes are independent conditional on the observed covariates and an unmeasured latent variable. Under this assumption, we prove the identification of the TBR and THR in non-separable (generalized) linear models for both continuous and binary outcomes. We then propose estimators and derive their asymptotic distributions. In the simulation studies, we implement our proposed methods to assess the performance of our estimators and carry out a sensitive analysis for different underlying distribution for the latent variable. Finally, we illustrate the proposed methods in two randomized controlled trials.
\end{Abstract}
\vfill
\end{titlepage}
\def\thefootnote{\fnsymbol{footnote}}

\section{Introduction}
\begin{spacing}{1.2}

The average treatment effect (ATE) is popularly used in evaluating the effect of a treatment or intervention in a wide range of disciplines such as medicine, social sciences, econometrics and etc. An assumption implicitly made by the ATE is the similarity of treatment effect across heterogeneous individuals. Although this assumption maybe warranted for some treatments, it is less plausible for others. For example, most patients treated with MMR (measles, mumps, and rubella) vaccine benefit from a very low risk of having Measles (one dose of MMR vaccine is about 93\% effective while two doses are about 97\% effective at preventing measles if exposed to the virus). In contrast, clinical evidence was found that prescription of a beta-blocker may or may not provide the desired response in treating patients with hypertension \cite{bradley2007beta}. Likewise, the prescription of anti-anxiety drugs such as Benzodiazepines may or may not be effective in treating patients with anxiety: some patients suffer from side effects such as drowsiness and depression while some others experienced paradoxical reactions such as increased anxiety, irritability, and agitation.

Formally, the heterogeneity of treatment effect is present if the effect of the treatment varies across subsets of individuals in a population \cite{poulson2012treatment}. This variability at the individual level is also called subject-treatment interaction \cite{gadbury2001evaluating, gadbury2004individual}. The heterogeneity of treatment effect may not only arise from different baseline characteristics of individuals (also known as pre-treatment heterogeneity) such as age, sex, social status but also from distinct individual responses to a particular treatment or intervention \cite{brand2013causal}.

The importance of treatment effect heterogeneity has been frequently raised by practical \mbox{concerns}. From a clinician's perspective, the heterogeneity of treatment effect plays an essential role in selecting the most effective treatment and designing individualized treatment regimens \cite{imai2013estimating}. Additionally, it is critical for policy makers to understand the heterogeneity effect so as to generalize causal effect estimates obtained from an experimental sample to a target population.

With the observed effect modifiers, the conditional ATE for different subpopulations is typically calculated.  In principle, such subgroup analysis would yield homogeneous treatment effect controlling for all effect modifications. However, it is hard to target and collect all effect modifiers based on the existing knowledge and limited resources \cite{zhang2013assessing}. As a result, the residual heterogeneity stand in the way of better understanding the treatment effect and more effectively designing the optimal treatment for each individual. Furthermore, the evidence of heterogeneous treatment effect urges further pursuit of unknown effect modifiers. Novel methods are thus of demand to assess the treatment effect heterogeneity of the study population or subpopulation.

To better illustrate the treatment effect heterogeneity, we use the framework of potential outcomes \cite{rubin1974estimating, rosenbaum1983central, holland1986statistics}. Under this framework, each individual has a potential outcome for every possible treatment, and the individual level effect of an experimental treatment relative to a control is defined by a comparison between the corresponding potential outcomes. However, for each individual, only one potential outcome, the one corresponds to the actual treatment, can be observed in practice.

Under the potential outcomes framework, the treatment benefit rate (TBR) and the treatment harm rate (THR) have been defined to assess the treatment effect heterogeneity \cite{gadbury2000unit, gadbury2001evaluating, gadbury2004individual, albert2005assessing, poulson2012treatment, shen2013treatment, zhang2013assessing}. More specifically, when the outcomes are binary, the TBR (THR) is defined as the rate of people who have a better (worse) outcome if given the treatment compared with control. The TBR and THR can be similarly defined for continuous outcomes by comparing the difference between the potential outcomes with some level $c$. Note that the definitions of the TBR and THR involve the joint distribution of the two potential outcomes, thus can not be identified without further assumptions even in randomized trials.

There is a growing number of literature deriving the bounds for the TBR and THR. Gadbury {\it et al.} \cite{gadbury2004individual} derived the simple bounds of the THR by only using the observed data without further assumptions. Additionally, they derived tighter bounds by estimating the quality of matching in a matched design. Albert {\it et al.} \cite{albert2005assessing} extended the method to a block trial which includes the matched trial as a special case. Under the assumption that subjects are exchangeable within blocks and that the within-block probabilities are constant across blocks, they estimated the bounds and provided the variances for the estimators. Alternatively, Yin and Zhou \cite{yin2016using} used a secondary outcome to obtain tighter bounds under the monotonicity, transitivity and causal necessity assumptions.

Other attempts have also been made to identify and estimate the TBR and THR under independence assumptions. Shen {\it et al.} \cite{shen2013treatment} and Zhang {\it et al.} \cite{zhang2013assessing} assumed that the two potential outcomes were independent conditional on observed covariates. However, this assumption is stringent in practice since the two potential outcomes are from the same individual and there is no guarantee that all the observed covariates are sufficient to explain the dependence. Yin {\it et al.} \cite{yin2016assessing} estimated the TBR and THR assuming the existence of at least three covariates which are mutually independent in the subgroup defined by the joint distribution of the potential outcomes. Their assumption could be tested when more than three such covariates were available without any modeling assumptions. However, it is hard to find such covariates in practice and extend the method to the case with continuous outcomes.

In this article, we assume that the potential outcomes are independent given the observed covariates and an unmeasured latent variable. This assumption is weaker than the independence assumption made by Shen {\it et al.} \cite{shen2013treatment} and Zhang {\it et al.} \cite{zhang2013assessing} and is much more reasonable in practice. Under non-separable (generalized) linear models, we prove the identification and construct estimators using maximum-likelihood estimation (MLE). We also derive the asymptotic distribution and variance for the estimators.

We organize the paper as follows. In Section 2, we introduce the notations and describe the assumptions. In Section 3, we provide identification conditions for the TBR and THR under non-separable models for continuous and binary outcomes. The estimators and their asymptotic properties are derived in Section 4. We report the simulation results in Section 5. Then we illustrate our proposed method in two randomized trials in Section 6. The paper concludes with a discussion in Section 7.

\section{Preliminaries}
Let $T$ denote a binary treatment assignment variable which is completely randomized and let $Y$ denote a primary outcome of interest. Let $X=(X_{1},\cdots,X_{p})^{T}$ denote p-dimensional observed covariates, where the superscript $^{T}$ denotes transposition. Let $t$ denote a possible value $T$ could take ($t=1$ for treatment and $t=0$ for placebo). Assume larger value of $Y$ indicates better response. Under the Stable Unite Treatment Value Assumption (SUTVA) \cite{rubin1980randomization}, we denote $Y_{1}$ and $Y_{0}$ as the potential outcomes under treatment and control, respectively.

When the outcome variable $Y$ is binary, Shen {\it et al.} \cite{shen2013treatment} defined the TBR and THR as
$$\mathrm{TBR}=P(Y_{0}=0,Y_{1}=1)~\mathrm{and}~\mathrm{THR}=P(Y_{0}=1,Y_{1}=0).$$
We can also define the TBR (THR) for subpopulation with specific covariates $X$ as $\mathrm{TBR}(X)=P(Y_{0}=0,Y_{1}=1|X)$. For the simplicity of illustration, we only focus on the population TBR and THR here. The inference for the TBR and THR in the subpopulation could be derived in a similar fashion. Note the TBR is the proportion of individuals in the population that have a better outcome if given treatment compared to control. In contrast, the THR is the proportion of individuals in the population that have a better outcome if given control compared to treatment. Note that when the outcomes are binary, the ATE corresponds to the difference between the TBR and the THR, which not only provide information about the overall treatment effect but also how treatment effect may vary across individuals. When $Y$ is continuous, we extend the definition of the TBR and the THR by comparing the difference between the potential outcomes with some level $c$¡±. More specifically, define:
$$\mathrm{TBR}_{c}=P(Y_{1}-Y_{0}>c)~\mathrm{and}~\mathrm{THR}_{c}=P(Y_{0}-Y_{1}>c),$$
where $c$ is a pre-specified constant. Again, $\mathrm{TBR}_{c}$ and $\mathrm{THR}_{c}$ can be defined analogously for subpopulation with specific covariates values, e.g., define $\mathrm{TBR}_{c}(X)=P(Y_{1}-Y_{0}<c|X)$. Note that $\mathrm{TBR}_{c}$ is the proportion of individuals in the population whose outcome $Y$ would benefit greater than $c$ from the treatment compared with the control and $\mathrm{THR}_{c}$ is the proportion of the individuals whose outcome $Y$ would be harmed by at least $c$ by the treatment compared with the control.

%
%

Due to the randomization, we can identify the marginal distributions of $Y_{0},Y_{1}$ as well as $\mathrm{ATE}=E(Y_{1}-Y_{0})$. However, as mentioned previously, the TBR, THR, $\mathrm{TBR}_{c}$ and $\mathrm{THR}_{c}$ involve the joint distribution of the two potential outcomes, thus can not be identified even in randomized trails without any further assumption. To make progress, Shen {\it et al.} \cite{shen2013treatment} and Zhang {\it et al.} \cite{zhang2013assessing} made the following assumption.

\textbf{Assumption 1.} (Conditional Independence) $Y_{0}\bot Y_{1}|X.$

Assumption 1 states that the two potential outcomes are independent conditional on a set of observed relevant baseline covariates.
Hence, the joint distribution of $Y_{0}$ and $Y_{1}$ can be identified by factorization, i.e., $P(Y_{0},Y_{1}|X)=P(Y_{0}|X)P(Y_{1}|X)$. However, this assumption requires the collection of all relevant covariates $X$ to control for the dependency between two potential outcomes, which is hard to satisfy in practice and impossible to test from the observed data. Alternatively, we make the following assumption by assuming the independence between the potential outcomes conditional on observed covariates $X$ as well as a latent variable $U$.

\textbf{Assumption 2.} (Latent Independence) $Y_{0}\bot Y_{1}|(X,U), ~~U\bot X.$

Note that Assumption 1 is a special case of Assumption 2 when there is no latent variable $U$, i.e., $U\in \emptyset$.
Also note that the independence between $X$ and $U$ can be satisfied by orthogonalization of  $U$ with respect to $X$ \cite{zhang2013assessing}. Additionally, Zhang {\it et al.} \cite{zhang2013assessing} claimed that under Assumption 2, the information of $U$ is not identifiable in a generalized linear mixed model (GLMM) and thus adopted a sensitivity analysis.


\section{Identification}

In this section, we derive the identification for the TBR and the THR under non-separable GLMM for both continuous and binary outcomes. More specifically, we have the following model for continuous outcomes:
\begin{equation}\label{Model continuous}
\left\{
  \begin{array}{ll}
Y_{t}=\alpha_{t,0}+\alpha_{t,1}^{T}X+\alpha_{t,2}U+\alpha_{t,3}^{T}XU+\epsilon_{t},\\
\epsilon_{t}\bot(X,U),\epsilon_{t}\sim N(0,\sigma_{t}^{2}),U\sim N(\mu_{U},\sigma^{2}_{U}), \alpha_{t,3}\neq0,
\end{array}
\right.
\end{equation}
for $t=0,1$, where $\alpha_{t,1}=(\alpha_{t,1}^{(1)},\cdots,\alpha_{t,1}^{(p)})^{T},\alpha_{t,3}=(\alpha_{t,3}^{(1)},\cdots,\alpha_{t,3}^{(p)})^{T}$. Without loss of generality, we can assume $\alpha_{t,2}>0$ since otherwise set $U^{*}=\mathrm{sign}(\alpha_{t,2})\cdot U$ and $\alpha_{t,2}^{*}=\mathrm{sign}(\alpha_{t,2})\cdot \alpha_{t,2}$, where $\mathrm{sign}(k)$ denotes the sign of $k$. Note that $\alpha_{t,3}\neq 0$ indicates that the model is not separable, i.e., the model can not be written in the form of $Y_t=l_1(X)+l_2(U)$. This is not a very stringent assumption especially when the observed covariates $X$ is high dimensional since we only require at least one but not all interactions between $X$ and $U$. Especially the existence of the interaction can be tested by checking whether $\alpha_{t,3}$ is significant with the observed data, although $U$ is a latent variable. Note that the latent variable $U$ can be interpreted as a subject specific random effect and the distribution of $U$ is assumed to be normal distribution. We will test how the identification and estimation method perform when the normality assumption of $U$ is violated in a sensitivity analysis in Section 5. Also without loss of generality, we can assume $(\mu_{U},\sigma^{2}_{U})=(0,1)$ since otherwise $U$ can be standardized.

Note that the inclusion of $U$ in the model (\ref{Model continuous}) can be tested by checking whether that the coefficients $(\alpha_{t,2},\alpha_{t,3},t=0,1)$ are significant. If there are at least one of them is significant in the models for both $Y_{0}$ and $Y_{1}$, then the Assumption 1 is violated and we must include a latent $U$ to make the conditional independence of $Y_{0}$ and $Y_{1}$ to hold. 

Note that we have the following formulas (\ref{Formula1})-(\ref{Formula2}) for $\mathrm{TBR}_{c}$ and $\mathrm{THR}_{c}$, the proofs of which are given in the supplementary material. From (\ref{Formula1})-(\ref{Formula2}), we know that once the parameters in model (\ref{Model continuous}) are identified, the $\mathrm{TBR}_{c}$ and $\mathrm{THR}_{c}$ can also be identified. Specifically,
\begin{equation}\label{Formula1}\tag{F1}
\mathrm{TBR}_{c}=\int\Phi\Big(\frac{(\alpha_{1,0}-\alpha_{0,0})+(\alpha_{1,1}-\alpha_{0,1})^{T}x-c}
{\sqrt{\big((\alpha_{1,2}-\alpha_{0,2})+(\alpha_{1,3}-\alpha_{0,3})^{T}x\big)^{2}+\sigma_{0}^{2}+\sigma_{1}^{2}}}\Big)f_{X}(x)dx,
\end{equation}
\begin{equation}\label{Formula2}\tag{F2}
\mathrm{THR}_{c}=\int\Phi\Big(\frac{(\alpha_{0,0}-\alpha_{1,0})+(\alpha_{0,1}-\alpha_{1,1})^{T}x-c}
{\sqrt{\big((\alpha_{0,2}-\alpha_{1,2})+(\alpha_{0,3}-\alpha_{1,3})^{T}x\big)^{2}+\sigma_{0}^{2}+\sigma_{1}^{2}}}\Big)f_{X}(x)dx,
\end{equation}
where $f_{X}(\cdot)$ is the density of $X$, $\Phi(\cdot)$ is the cumulative distribution function of a standard normal variable.

When outcomes are binary, we consider the following model:
\begin{equation}\label{Model binary}
\left\{
  \begin{array}{ll}
Y_{t}^{*}=\alpha_{t,0}+\alpha_{t,1}^{T}X+\alpha_{t,2}U+\alpha_{t,3}^{T}XU+\epsilon_{t},\\
Y_{t}=I(Y_{t}^{*}>0),\\
\epsilon_{t}\bot(X,U),\epsilon_{t}\sim N(0,\sigma_{t}^{2}),U\sim N(\mu_{U},\sigma^{2}_{U}), (\alpha_{t,0},\alpha_{t,1})\neq0, \alpha_{t,3}\neq0,
\end{array}
\right.
\end{equation}
for $t=0,1$, where $\alpha_{t,1}=(\alpha_{t,1}^{(1)},\cdots,\alpha_{t,1}^{(p)})^{T},\alpha_{t,3}=(\alpha_{t,3}^{(1)},\cdots,\alpha_{t,3}^{(p)})^{T}$. Again, we assume that $U$ follows the standard normal distribution. Additionally, without loss of generality, we can assume $\sigma^{2}_{0}=\sigma^{2}_{1}=1$ since otherwise set $Y^{**}_{t}=Y^{*}_{t}/\sigma_{t},\alpha_{t,k}=\alpha_{t,k}/\sigma_{t},$ and $\epsilon_{t}^{*}=\epsilon_{t}/\sigma_{t}$ for $t=0,1$ and $k=0,1,2,3.$
Note that $Y^{*}$ is a latent variable and (\ref{Model binary}) indicates a probit model for the outcome $Y$, i.e.,
$$P(Y_{t}=1|X,U)=\Phi\Big(\alpha_{t,0}+\alpha_{t,1}^{T}X+\alpha_{t,2}U+\alpha_{t,3}^{T}XU\Big).$$

Also, similar to the continuous model (\ref{Model continuous}), the inclusion of $U$ and the interaction terms can be tested by checking corresponding coefficients are significant.

Note that we have the following formulas (\ref{Formula3})-(\ref{Formula4}) for TBR and THR, the proofs of which are given in the supplementary material. From (\ref{Formula3})-(\ref{Formula4}), we know that once the parameters in model (\ref{Model binary}) are identified, the TBR and THR can also be identified. Specifically,
\begin{equation}\label{Formula3}\tag{F3}
\mathrm{TBR}=\int \Phi_{b}\big(\mu(x;\theta),\Sigma(x;\theta)\big)f_{X}(x)dx,
\end{equation}
\noindent and
\begin{equation}\label{Formula4}\tag{F4}
\mathrm{THR}=\int \Phi_{h}\big(\mu(x;\theta),\Sigma(x;\theta)\big)f_{X}(x)dx,
\end{equation}
where
$$\mu(x;\theta)=\big(\mu_{0}(x;\theta),\mu_{1}(x;\theta)\big)=\big(-\alpha_{0,0}-\alpha_{0,1}^{T}x,~-\alpha_{1,0}-\alpha_{1,1}^{T}x\big),$$
$$\Sigma(x;\theta)=\left(
                \begin{array}{cc}
                  1+(\alpha_{0,2}+\alpha_{0,3}^{T}x)^{2}  & (\alpha_{0,2}+\alpha_{0,3}^{T}x)(\alpha_{1,2}+\alpha_{1,3}^{T}x) \\
                  (\alpha_{0,2}+\alpha_{0,3}^{T}x)(\alpha_{1,2}+\alpha_{1,3}^{T}x) & 1+(\alpha_{1,2}+\alpha_{1,3}^{T}x)^{2} \\
                \end{array}
              \right).$$
$$\Phi_{b}\big(\mu,\Sigma\big)= \Phi_{2}\big((0,\infty),(-\infty,0);\mu,\Sigma\big),$$
$$\Phi_{h}\big(\mu,\Sigma\big)= \Phi_{2}\big((-\infty,0),(0,\infty);\mu,\Sigma\big),$$
and $\Phi_{2}(A_{0},A_{1};\mu,\Sigma)$ is the distribution function of bivariate normal vector with mean $\mu$, covariance matrix $\Sigma$ in the integration region $A_{0}\times A_{1}$, i.e.,
$$\Phi_{2}(A_{0},A_{1};\mu,\Sigma)=\int\int_{A_{0}\times A_{1}}\frac{1}{2\pi|\Sigma|^{1/2}}\exp\big\{-\frac{1}{2}(s_{0}-\mu_{0},s_{1}-\mu_{1})\Sigma^{-1}(s_{0}-\mu_{0},s_{1}-\mu_{1})^{T}\big\}ds_{0}ds_{1}.$$

Let $\theta=(\theta_{0}^{T},\theta_{1}^{T})^{T}$ denote the parameters in models (\ref{Model continuous}) and (\ref{Model binary}). More specifically, let $\theta_{t}=(\alpha_{t,0},\alpha_{t,1}^{T},\alpha_{t,2},\alpha_{t,3}^{T},\sigma_{t}^{2})^{T}$ in the continuous model and $\theta_{t}=(\alpha_{t,0},\alpha_{t,1}^{T},\alpha_{t,2},\alpha_{t,3}^{T})^{T}$ in the binary model, for $t=0,1$. We can have the following theorem for the identification of $\theta$ and thus the identification of $(\mathrm{TBR}_{c},~\mathrm{THR}_{c})$ for the continuous outcomes and (TBR, THR) for the binary outcomes, the proof of which is given in the supplementary material.

\textbf{Theorem 1.}  Under Assumption 2,
\begin{enumerate}
  \item [(\romannumeral1)] If the model (\ref{Model continuous}) holds for continuous outcomes, the parameters $\theta$ can be identified, thus the $\mathrm{TBR}_{c}$ and $\mathrm{THR}_{c}$ can also be identified for any constant $c$.
  \item [(\romannumeral2)] If the model (\ref{Model binary}) holds for binary outcomes and the regularity Condition C. (given in the supplementary material) holds, the parameters $\theta$ can be identified, thus the TBR and THR can also be identified.
\end{enumerate}

Theorem 1 states the identification of GLMM for both continuous and binary outcomes under a relaxed conditionally independence of potential outcomes conditional on observed covariates and a latent variable $U$. Note that the non-separable condition plays an important role in the identification of the parameters $\theta$ in the presence of $U$. When marginalizing over $U$, the interaction term between $U$ and $X$ helps identify the effect of $U$ on $Y$ which would otherwise be absorbed in intercept.

Note that the covariates $X$ are in their linear terms in model (1) and (2). However, this is not required for the identification purpose. When the outcomes are continuous, the general form of the GLMM is

\begin{equation}\label{general_con}
\left\{
  \begin{array}{ll}
Y_{t}=g_{t}(X)+Uh_{t}(X)+\epsilon_{t},\\
\epsilon_{t}\bot(X,U),\epsilon_{t}\sim N(0,\sigma_{t}^{2}),U\sim N(0,1),
\end{array}
\right.
\end{equation}
for $t=0,1$. Similarly, when the outcomes are binary, the general form  of a GLMM is
\begin{equation}\label{general_bin}
\left\{
  \begin{array}{ll}
Y_{t}^{*}=g_{t}(X)+Uh_{t}(X)+\epsilon_{t},\\
Y_{t}=I(Y_{t}^{*}>0),\\
\epsilon_{t}\bot(X,U),\epsilon_{t}\sim N(0,1),U\sim N(0,1),
\end{array}
\right.
\end{equation}
for $t=0,1$. Note that (\ref{Model continuous}) and (\ref{Model binary}) are special case of (\ref{general_con}) and (\ref{general_bin}) with $g_{t}(X)=\alpha_{t,0}+\alpha_{t,1}^{T}X$ and $h_{t}(X)=\alpha_{t,2}+\alpha_{t,3}^{T}X$. In the supplementary material, we give the necessary and sufficient conditions to identify $\theta=(g_{0}(X),h_{0}(X),\sigma^{2}_{0},g_{1}(X),h_{1}(X),\sigma^{2}_{1})$ in model (\ref{general_con}) and $\theta=(g_{0}(X),h_{0}(X),g_{1}(X),h_{1}(X))$ in model (\ref{general_bin}). Note that once $(h_{0}(X),h_{1}(X))$ can be identified, we can test the inclusion of $U$ by testing whether $(h_{0}(X),h_{1}(X))$ is significant with the observed data.

\section{Inference}
We now propose estimators for TBR and THR and derive their asymptotic properties.

\subsection{Continuous Outcomes}
Note that $\theta$ can be estimated by the MLE $\widehat{\theta}$ which is obtained by maximizing the log-likelihood,
$$\ell=\log L(T,X,Y)=P_{n}\big\{\psi(T,X,Y;\theta)\big\},$$
where $P_{n}$ is the empirical mean, i.e., $P_{n}g(X)=\sum_{i=1}^{n}g(X_{i})/n$, and
\begin{eqnarray*}
 &&\psi(T,X,Y;\theta)\\
  &=& \sum_{t=0,1}\frac{1}{2}\bigg[I(T=t)\Big\{-\log(2\pi)-\log\big((\alpha_{t,2}+\alpha_{t,3}^{T}X)^{2}+\sigma_{t}^{2}\big)
-\frac{(Y-\alpha_{t,0}-\alpha_{t,1}^{T}X)^{2}}{(\alpha_{t,2}+\alpha_{t,3}^{T}X)^{2}+\sigma_{t}^{2}}\Big\}\bigg].
\end{eqnarray*}
Thus we have the asymptotic normality of $\sqrt{n}(\widehat{\theta}-\theta)$, which can be used to test the significant of the parameters.
Following (\ref{Formula1}) and (\ref{Formula2}), we can estimate the $\mathrm{TBR}_{c}$ and $\mathrm{THR}_{c}$ by

$$\widehat{\mathrm{TBR}}_{c}=P_{n}\Phi\Big(\frac{(\widehat{\alpha}_{1,0}-\widehat{\alpha}_{0,0})+(\widehat{\alpha}_{1,1}-\widehat{\alpha}_{0,1})^{T}X-c}
{\sqrt{\big((\widehat{\alpha}_{1,2}-\widehat{\alpha}_{0,2})+(\widehat{\alpha}_{1,3}-\widehat{\alpha}_{0,3})^{T}X\big)^{2}
+\widehat{\sigma}_{0}^{2}+\widehat{\sigma}_{1}^{2}}}\Big),$$
$$\widehat{\mathrm{THR}}_{c}=P_{n}\Phi\Big(\frac{(\widehat{\alpha}_{0,0}-\widehat{\alpha}_{1,0})+(\widehat{\alpha}_{0,1}-\widehat{\alpha}_{1,1})^{T}X-c}
{\sqrt{\big((\widehat{\alpha}_{0,2}-\widehat{\alpha}_{1,2})+(\widehat{\alpha}_{0,3}-\widehat{\alpha}_{1,3})^{T}X\big)^{2}
+\widehat{\sigma}_{0}^{2}+\widehat{\sigma}_{1}^{2}}}\Big),$$

where $\widehat{\alpha}_{t,k}$ and $\widehat{\sigma}_{t^2}$ are MLEs for the corresponding parameters.
The following theorem shows the $\sqrt{n}$ consistency, asymptotic normality and provides the asymptotic variance of the estimators when the outcomes are continuous.

\textbf{Theorem 2.} If the model (\ref{Model continuous}) holds for continuous outcomes, we can have
$$\sqrt{n}\big(\widehat{\mathrm{TBR}}_{c}-\mathrm{TBR}_{c}\big)\stackrel{d}{\longrightarrow} N(0,\sigma_{cB}^{2}(\theta)),$$
$$\sqrt{n}\big(\widehat{\mathrm{THR}}_{c}-\mathrm{THR}_{c}\big)\stackrel{d}{\longrightarrow} N(0,\sigma_{cH}^{2}(\theta)),$$
\noindent where $\stackrel{d}{\longrightarrow}$ denotes convergence in distribution. The proof of Theorem 2 and the expressions and consistent estimators of $\sigma_{cB}^{2}(\theta)$ and $\sigma_{cH}^{2}(\theta)$ are given in the supplementary material.

\subsection{Binary Outcomes}
Note that $\theta$ can be estimated by the MLE $\widehat{\theta}$ which is obtained by maximizing the log-likelihood,
$$\ell=\log L(T,X,Y)=P_{n}\big\{\psi(T,X,Y;\theta)\big\},$$
and
\begin{eqnarray*}
 \psi(T,X,Y;\theta)  &=& \sum_{t=0,1}\bigg[I(T=t)\Big\{Y\log\big(G(X;\theta_{t})\big)+(1-Y)\log\big(1-G(X;\theta_{t})\big)\Big\} \bigg],
\end{eqnarray*}
where
$$G(X;\theta_{t})=\Phi\big(\frac{\alpha_{t,0}+\alpha_{t,1}^{T}X}{\sqrt{1+(\alpha_{t,2}+\alpha_{t,3}^{T}X)^{2}}}\big).$$
Similarly, we can have the asymptotic normality of $\sqrt{n}(\widehat{\theta}-\theta)$, which can be used to test the significant of the parameters.
Following (\ref{Formula3}) and (\ref{Formula4}), we can estimate TBR and THR by
$$\widehat{\mathrm{TBR}}=P_{n}\Phi_{b}\big(\mu(X;\widehat{\theta}),\Sigma(X;\widehat{\theta})\big),$$
$$\widehat{\mathrm{THR}}=P_{n}\Phi_{h}\big(\mu(X;\widehat{\theta}),\Sigma(X;\widehat{\theta})\big).$$

The following theorem shows the $\sqrt{n}$ consistency, asymptotic normality and provides the asymptotic variance of the estimators when the outcomes are binary.

\textbf{Theorem 3.} If the model (\ref{Model binary}) holds for binary outcomes and the regularity Condition C. holds, we have
$$\sqrt{n}\big(\widehat{\mathrm{TBR}}-\mathrm{TBR}\big)\stackrel{d}{\longrightarrow} N(0,\sigma_{bB}^{2}(\theta)),$$
$$\sqrt{n}\big(\widehat{\mathrm{THR}}-\mathrm{THR}\big)\stackrel{d}{\longrightarrow} N(0,\sigma_{bH}^{2}(\theta)).$$

The expressions and consistent estimators of $\sigma_{bB}^{2}(\theta)$ and $\sigma_{bH}^{2}(\theta)$ are given in the supplementary material.

\section{Simulation}
\subsection{The performance of the estimators}
We first assess the performance of the estimators proposed in Section 4. The simulations were conducted under two scenarios: (a) the continuous outcomes and (b) the binary outcomes. For scenario (a), the simulation study was conducted in the following steps:
\begin{enumerate}
  \item [Step 1:] A population of sample size 1000 was created. Variables $T$, $X$ and $U$ were generated independently. More specifically, treatment $T$ was generated from a Bernoulli distribution with $P(T=1)=0.5$, the components of covariates $X=(X_{1},X_{2},X_{3})^{T}$ were identically and independently generated from a standard normal distribution and latent variable $U$ was also generated from standard normal distribution. Potential outcomes $(Y_{0},Y_{1})$ were generated from model (\ref{Model continuous}) with the parameters set to:
$$(\alpha_{0,0},\alpha_{0,1}^{(1)},\alpha_{0,1}^{(2)},\alpha_{0,1}^{(3)},\alpha_{0,2},
\alpha_{0,3}^{(1)},\alpha_{0,3}^{(2)},\alpha_{0,3}^{(3)})=(-0.3,1.2,-1.0,-0.8,0.7,-0.5,1.3,0.6),$$
$$(\alpha_{1,0},\alpha_{1,1}^{(1)},\alpha_{1,1}^{(2)},\alpha_{1,1}^{(3)},\alpha_{1,2},
\alpha_{1,3}^{(1)},\alpha_{1,3}^{(2)},\alpha_{1,3}^{(3)})=(0.2,-0.8,1.2,1.0,0.8,-0.6,1.0,0.6),$$
$$\sigma_{0}^{2}=1.0,~\sigma_{1}^{2}=1.2.$$
  \item [Step 2:] The parameters $\theta$ was estimated using MLE and the estimates of $(\mathrm{TBR}_{c},\mathrm{THR}_{c})$ and the variances of the estimators were obtained, where $c=1$.
  \item [Step 3:] Steps 1 and 2 were repeated for 1000 times to obtain the biases, average estimated standard error (ASE) and the empirical standard error (ESE).
\end{enumerate}
The results are reported in Table \ref{simulation_con} where $U$ was simulated from a normal distribution. From the table we can see that the biases are -0.001 and 0.003 for $\mathrm{TBR}_{c}$ and $\mathrm{THR}_{c}$ respectively, and the ASE are 0.017 and 0.015 respectively, which both approximate their ESE. Also note that the coverages of the 95\% CI approximate 0.95 indicating the good performance of our estimators.

\begin{table}
\caption{The true value, bias, average estimated standard error (ASE), empirical standard error (ESE) and 95\% confidence interval (CI) coverage in continuous case. Every table cell contains two elements, which corresponds to $\mathrm{TBR}_{c}$ (first row in each cell) and $\mathrm{THR}_{c}$ (second row in each cell) ($c=1$) respectively.}\label{simulation_con}
\begin{center}
\begin{tabular}{cccccc}
  \hline
  Distribution of $U$& true value & bias  & ASE & ESE &  95\% CI coverage \\
  \hline
  \multirow{2}*{Normal}         & 0.450 & -0.001  & 0.017 & 0.017 & 0.935 \\
                                & 0.346 &  0.003  & 0.015 & 0.015 & 0.942 \\
  \hline
  \multirow{2}*{t(3)}           & 0.448 & -0.001  & 0.017 & 0.017 & 0.949 \\
                                & 0.347 &  0.004  & 0.015 & 0.016 & 0.935 \\
  \hline
  \multirow{2}*{t(10)}          & 0.448 & -0.000  & 0.017 & 0.017 & 0.951 \\
                                & 0.345 &  0.005  & 0.015 & 0.015 & 0.937 \\
  \hline
  \multirow{2}*{$\chi^{2}(3)$}  & 0.448 &  0.001  & 0.017 & 0.016 & 0.949 \\
                                & 0.349 &  0.002  & 0.015 & 0.015 & 0.953 \\
  \hline
  \multirow{2}*{$\chi^{2}(10)$} & 0.448 & -0.000  & 0.017 & 0.017 & 0.951 \\
                                & 0.346 &  0.005  & 0.015 & 0.015 & 0.938 \\
  \hline
  \multirow{2}*{P(3)}           & 0.449 &  0.000  & 0.017 & 0.017 & 0.950 \\
                                & 0.348 &  0.002  & 0.015 & 0.015 & 0.944 \\
  \hline
  \multirow{2}*{P(10)}          & 0.448 &  0.001  & 0.017 & 0.017 & 0.937 \\
                                & 0.345 &  0.004  & 0.015 & 0.015 & 0.928 \\
  \hline
  \multirow{2}*{B(0.5)}         & 0.450 & -0.000  & 0.017 & 0.017 & 0.955 \\
                                & 0.349 &  0.000  & 0.015 & 0.015 & 0.955 \\
  \hline
\end{tabular}
\end{center}
\end{table}

For the scenario (b), the simulation process is similar to that for scenario (a), except: in Step 1, we set the sample size to be 2000 and generate $(Y_{0},Y_{1})$ from model (\ref{Model binary}) with the same $\theta$ excluding $(\sigma_{0}^{2},\sigma_{1}^{2})$ and in Step 2, the (TBR, THR) were calculated instead of the $(\mathrm{TBR}_c, \mathrm{THR}_c$). The results for the binary outcomes are shown in Table \ref{simulation_bin} where $U$ is simulated from a normal distribution. From the table we can see that the biases are 0.001 and 0.001 for TBR and THR respectively, and the ASE are 0.016 and 0.013 respectively, which both approximate their ESE. Similarly, as in the scenario (a), the coverages of the 95\% CI approximate 0.95 indicating good performance of our estimators.

\begin{table}
\caption{The true value, bias, average estimated standard error (ASE), empirical standard error (ESE) and 95\% confidence interval (CI) coverage in continuous case. Every table cell contains two elements, which corresponds to TBR (first row in each cell) and THR (second row in each cell) respectively.}\label{simulation_bin}
\begin{center}
\begin{tabular}{cccccc}
  \hline
  Distribution & true value & bias &  ASE & ESE &  95\% coverage \\
  \hline
  \multirow{2}*{Normal}         & 0.363 &  0.001 & 0.016 & 0.016 & 0.947 \\
                                & 0.241 &  0.001 & 0.013 & 0.013 & 0.956 \\
  \hline
  \multirow{2}*{t(3)}           & 0.339 &  0.002 & 0.015 & 0.015 & 0.943 \\
                                & 0.305 &  0.001 & 0.013 & 0.014 & 0.948 \\
  \hline
  \multirow{2}*{t(10)}          & 0.306 &  0.001 & 0.016 & 0.015 & 0.954 \\
                                & 0.294 &  0.001 & 0.014 & 0.013 & 0.955 \\
  \hline
  \multirow{2}*{$\chi^{2}(3)$}  & 0.315 &  0.002 & 0.016 & 0.016 & 0.941 \\
                                & 0.301 &  0.002 & 0.014 & 0.014 & 0.945 \\
  \hline
  \multirow{2}*{$\chi^{2}(10)$} & 0.302 &  0.000  & 0.016 & 0.016 & 0.947 \\
                                & 0.297 &  0.000  & 0.014 & 0.014 & 0.949 \\
  \hline
  \multirow{2}*{P(3)}           & 0.297 &  0.000  & 0.016 & 0.016 & 0.951 \\
                                & 0.295 &  0.000  & 0.014 & 0.013 & 0.955 \\
  \hline
  \multirow{2}*{P(10)}          & 0.297 &  4e-04  & 0.016 & 0.016 & 0.954 \\
                                & 0.294 &  6e-04  & 0.014 & 0.013 & 0.954 \\
  \hline
  \multirow{2}*{B(0.5)}         & 0.243 &  0.004 & 0.017 & 0.016 & 0.941 \\
                                & 0.288 & -0.000 & 0.014 & 0.015 & 0.943 \\
  \hline
\end{tabular}
\end{center}
\end{table}

\subsection{Sensitive analysis with respect to the distribution of U}
We assumed that $U$ is normally distributed for the identification of the joint distribution of $(Y_{0},Y_{1})$. Since $U$ is unobserved, its actual distribution is unknown which may or may not be normal. We carry out a sensitivity analysis to evaluate the performance of the estimators for TBR and THR under different underlying distribution of $U$. We consider the following underlying distribution for $U$: t-distribution, chi-squared, Poisson and Bernoulli. The estimation was carried out the same as section 5.1 except $U$ is generated from the distributions above. Note that we standardized $U$ to have mean 0 and variance 1 under different distributions.

The results when the outcomes are continuous are shown in Table \ref{simulation_con}. When $U$ is generated from t(3), the biases are -0.001 and 0.004 for $\mathrm{TBR}_{c}$ and $\mathrm{THR}_{c}$ respectively, and the ASE are 0.017 and 0.015 respectively, which approximate to the ESE. Also, the coverages of 95\% CI are 0.949 and 0.935 respectively, which approximate 0.95. Similar performance was also observed when $U$ follows other distributions such as chi-squared, Poisson and Bernoulli. Note that as the degree of freedom increases in distributions such as chi-squared, Poisson, the standardized $U$ can be approximated by a normal distribution. Thus the good performances of estimators under these distributions were expected. But when the degree of freedom is small, the performance of estimators are robust for both symmetric distributions (e.g., t-distribution) and skewed distributions (e.g., chi-squared distribution). Moreover, the estimators are robust even for the discrete distribution (Poisson, Bernoulli). Also, the ASE approximate to the ESE under different distributions of $U$.

The results of the binary case are shown in Table \ref{simulation_bin}. When $U$ follows a t(3) distribution, the biases are 0.002 and 0.001 for TBR and THR respectively, and the ASE are 0.015 and 0.013 respectively, which approximate to the ESE. Also, the coverages are 0.943 and 0.948 respectively, which approximate 0.95. From the table we can also conclude that when the outcomes are binary, the estimators are still robust to the different distribution of unmeasured variable $U$ including symmetric, non-symmetric and discrete distribution.

\section{Real data analysis}
We illustrate our methods in two randomized controlled trials.
\subsection{The Effect of Arnica 30$\times$ in Alleviating DOMS}
The delayed-onset muscle soreness (DOMS) is commonly experienced muscle tenderness or debilitating pain after exercising. The soreness usually reaches its peak in 24-48 hours after exercise. Despite the prevalence of DOMS, it still remains unclear about the mechanisms, treatment, and how it affects the athletic performance \cite{cheung2003delayed}.

A randomized, double-blinded placebo-controlled trial was carried out to determine whether homeopathic Arnica 30$\times$ can be beneficial for DOMS following long-distance running \cite{vickers1998homeopathic}. A total of 519 runners were randomized to either Arnica 30$\times$ or an indistinguishable placebo. The data results were obtained from 400 individuals while the rest 119 individuals did not run or were lost to follow up. Despite the missing data, the remaining 400 subjects were still considered to be randomized due to the double-blind design and well balanced baseline covariates distributions for treatment and control groups. The primary outcome measure was mean 11-point visual analog score (VAS) for the 2-day period after the run. The VAS score is continuous and ranges from 0 to 100, where the higher the VAS score indicates more muscle soreness. It has been found that the mean VAS score is 4.15 mm lower in the placebo group (95\% CI: (-0.51, 8.81)). Note that 0 is contained in the CI thus suggesting that that there is no significant effect of Arnica 30$\times$ in alleviating the DOMS.

Usually in a study with a non-significant ATE, there is still concern that if the treatment is beneficial to some individuals, we may still need to make such treatment as an option available for these who benefits. Let $Y=100-\text{VAS}$ to guarantee that larger value of $Y$ indicates better response.
Let ``age", ``sex", ``tr", ``le", ``inj" and ``rt" denote the age, gender, training miles, length of race, whether injured and race time of the individuals.
Since these six variables are all possibly related to the DOMS, we include them in model (\ref{Model continuous}). The results of the regression suggest some significant interactions, including U$\cdot$ inj ($\widehat{\alpha}_{0,3}^{(5)}=-6.58, P=0.004$, 95\% CI: (-11.50, -1.66)), U$\cdot$ tr ($\widehat{\alpha}_{1,3}^{(3)}=0.18, P=0.008$, 95\% CI: (0.04, 0.32)), U$\cdot$ le ($\widehat{\alpha}_{1,3}^{(4)}=-0.72, P=0.025$, 95\% CI: (-1.43,-0.02)), U$\cdot$ rt ($\widehat{\alpha}_{1,3}^{(6)}=0.06, P=0.030$, 95\% CI: (0.01, 0.12)). This justifies the inclusion of $U$ in the model (\ref{Model continuous}) and the non-separable assumption holds. We then estimated the $\mathrm{TBR}_{c}$ and $\mathrm{THR}_{c}$ over a range of pre-specified constant $c$.  The results are shown in Table \ref{Real data continuous}.
\begin{table}
\caption{Estimates and 95\% confidence intervals for $\mathrm{TBR}_{c}$ and $\mathrm{THR}_{c}$ with different values of pre-specified constant c for the study of treatment effect of Arnica 30$\times$ on DOMS}\label{Real data continuous}
\begin{center}
\begin{tabular}{ccccccc}
  \hline
   & \multicolumn{2}{c}{$\mathrm{TBR}_{c}$} & &\multicolumn{2}{c}{$\mathrm{THR}_{c}$} \\
\hline
  c & Estimate  & 95\% CI & &Estimate  & 95\% CI \\
  0 & 0.30 &  (0.00, 0.67) & & 0.70 & (0.33, 1.00)\\
  2 & 0.19 &  (0.00, 0.49) & & 0.55 & (0.17, 0.92)\\
  4 & 0.11 &  (0.00, 0.33) & & 0.39 & (0.04, 0.74)\\
  6 & 0.07 &  (0.00, 0.20) & & 0.25 & (0.00, 0.59)\\
  8 & 0.04 &  (0.00, 0.13) & & 0.16 & (0.00, 0.44)\\
 10 & 0.02 &  (0.00, 0.08) & & 0.10 & (0.00, 0.28)\\
 12 & 0.01 &  (0.00, 0.05) & & 0.07 & (0.00, 0.18)\\
 14 & 0.01 &  (0.00, 0.03) & & 0.05 & (0.00, 0.14)\\
 16 & 0.01 &  (0.00, 0.02) & & 0.04 & (0.00, 0.10)\\
 18 & 0.00 &  (0.00, 0.01) & & 0.03 & (0.00, 0.08)\\
 20 & 0.00 &  (0.00, 0.01) & & 0.02 & (0.00, 0.06)\\
  \hline
\end{tabular}
\end{center}
\end{table}

Note that the 95\% confidence intervals of $\mathrm{TBR}_{c}$ always contain 0, which means the Arnica 30$\times$ may not be beneficial to anyone. Note when $c=0$, we have $\widehat{\mathrm{THR}}_{c}=0.70$ with a 95\% confidence interval (0.33, 1.00) excluding 0. This indicates that at least 33\% of individual have worse outcome if given Arnica 30$\times$ as compared with control. Such harmful effect rate decreases as $c$ increase and goes away when we set $c\geq6$. In this study, since there is no subgroup of individual that might benefit from Arnica 30$\times$ and some portion of individuals that might have slightly more soreness using it. Thus, we reach the conclusion that there is no significant evidence support the use of using Arnica 30$\times$ to alleviate DOMS.

\subsection{ACCORD Eye Study}
The diabetic retinopathy (DR) is one of the most common causes of vision loss among people with diabetes and the leading cause of vision impairment and blindness among adults younger than 65 years old in the United States \cite{chew2007rationale}. It has been found that poor glycemic control is one of the most important risk factors associated with the development of DR. The Action to Control Cardiovascular Risk in Diabetes (ACCORD) study group enrolled 10,251 participants with type 2 diabetes who were at high risk for cardiovascular disease to randomly receive either intensive or standard treatment for glycemia randomly (target glycated hemoglobin level, $<$ 6.0\% or 7.0 to 7.9\%, respectively \cite{doi:10.1056/NEJMoa1001288}. Among those participants, there were 2856 of them were eligible for the ACCORD Eye study. The ACCORD Eye study aimed at determining whether the intensive glycemia could reduce the risk of development or progression of diabetic retinopathy, as compared with the standard treatments. The primary outcome of this study was the composite end point of either progression of diabetic retinopathy by at least three steps on the Early Treatment Diabetic Retinopathy Study (ETDRS) Severity Scale or development of proliferative diabetic retinopathy necessitating photocoagulation therapy or vitrectomy in 4 years \cite{doi:10.1056/NEJMoa1001288}. Since the intervention in the original study is randomized and the inclusion criteria did not affect the intervention, the ACCORD Eye study can still be considered as a randomized trial.

At the end of 4 years of follow-up, progression of diabetic retinopathy was seen in 7.3\% of participants (104 out of 1429) in the intensive glycemic control group, as compared with 10.4\% of participants (149 out of 1427) in the standard glycemic therapy group. Using the same notation as in Section 2, we denote $Y=0$ if the progression of diabetic retinopathy was seen, and $Y=1$ otherwise. Thus the ATE$=0.032$ (P = 0.003, 95\% CI, 0.011 to 0.052) suggesting a significant effect of the intensive glycemic control.

Usually in a study with significant ATE, there is still concern that the active treatment might be harmful for some individuals, thus we may need to proceed with caution when providing them with such intervention. To estimate the TBR and THR, we adjust in model (2) for other treatments that patients are using with indicator variables, denote as $(X_{1},X_{2},X_{3},X_{4})$. This set of covariates were also considered by Group and Group \cite{doi:10.1056/NEJMoa1001288} in their models.
The results of the regression suggest some significant interactions, including U$\cdot X_{1}$ ($\widehat{\alpha}_{0,3}^{(1)}=0.56, P=0.021$, 95\% CI: (0.24, 0.87)), U$\cdot X_{2}$ ($\widehat{\alpha}_{1,3}^{(2)}=0.53, P=0.027$, 95\% CI: (0.21, 0.85)). This justifies the inclusion of $U$ in the model (\ref{Model binary}) and the non-separable assumption holds.
The estimates for TBR and THR are 0.078 and 0.046 with their standard errors 0.006 and 0.005 respectively (p-value both $<0.0001$). Thus it is suggested that there are 7.8\% of people benefit from the intensive glycemia control while 4.6\% of people get harmed from intensive glycemia control.

\section{Discussion}

In this article, we assessed the treatment effect heterogeneity by evaluating the TBR and the THR. We relaxed the conditional independence Assumption 1 by allowing the presence of an unmeasured latent variable. Under our non-separable (generalized) linear models, the existence of the latent variable can be tested, and then we provided identification and estimation methods. The models we used require an interaction term between the latent variable and at least one covariate, which is likely to hold when covariates $X$ is high dimensional and can also be tested by the observed data. It can be shown that the parameters are not identifiable in the absence of such interaction due to the presence of unmeasured latent variable.

We imposed a normality assumption of the latent variable $U$. The normality of $U$ is not necessary for identification purpose, but when the distribution of $U$ is not normal, the distribution of $Y-g(X)$ conditional on $X$ may not have a distribution in closed form and the identification condition may thus be complicated.  We carried out a sensitivity analysis to evaluate the performance of estimators under different underlying distribution of $U$. We leave the generalization of identification and estimation of treatment effect heterogeneity under different distribution of $U$ as future research topic.

In our simulation studies, the estimation procedures are almost convergent for the cases of continuous outcomes, although they may not be so for a few cases of binary outcomes. For example, when the sample size is 2000, for binary outcomes, when the underlying distributions of $U$ is actually normal, we observed 0.9\% of non-convergence and when we do sensitivity analysis of $U$'s distribution, the non-convergence is 5.8\% for chi-squared distribution. Such non-convergence disappear when sample size is increased. Also, there was no such non-convergence observed for continuous outcomes in the simulation.

\Acknowledgements
The content is solely the responsibility of the authors. The authors thank Professor Lan Wang for helpful discussion and Professor Andrew J. Vickers for his generosity of making data available from the Arnica 30$\times$ experiment described in Section 6.



\end{spacing}

\bibliographystyle{plain}

\bibliography{TBR_arx}

\begin{thebibliography}{10}

\bibitem{albert2005assessing}
Jeffrey~M Albert, Gary~L Gadbury, and Edward~J Mascha.
\newblock Assessing treatment effect heterogeneity in clinical trials with
  blocked binary outcomes.
\newblock {\em Biometrical Journal}, 47(5):662--673, 2005.

\bibitem{bradley2007beta}
HA~Bradley, BM~Mayosi, RT~Maroney, A~Mbewu, L~Opie, and J~Volmink.
\newblock Beta-blockers for hypertension.
\newblock {\em Cochrane Database of Systematic Reviews}, 24:CD002003, 2007.

\bibitem{brand2013causal}
Jennie~E Brand and Juli~Simon Thomas.
\newblock Causal effect heterogeneity.
\newblock In {\em Handbook of Causal Analysis for Social Research}, pages
  189--213. Springer, 2013.

\bibitem{cheung2003delayed}
Karoline Cheung, Patria~A Hume, and Linda Maxwell.
\newblock Delayed onset muscle soreness.
\newblock {\em Sports Medicine}, 33(2):145--164, 2003.

\bibitem{chew2007rationale}
Emily~Y Chew, Walter~T Ambrosius, Letitia~T Howard, Craig~M Greven, Samantha
  Johnson, Ronald~P Danis, Matthew~D Davis, Saul Genuth, Michael Domanski,
  ACCORD~Study Group, et~al.
\newblock Rationale, design, and methods of the action to control
  cardiovascular risk in diabetes eye study (accord-eye).
\newblock {\em The American Journal of Cardiology}, 99(12):S103--S111, 2007.

\bibitem{gadbury2000unit}
Gary~L Gadbury and Hari~K Iyer.
\newblock Unit--treatment interaction and its practical consequences.
\newblock {\em Biometrics}, 56(3):882--885, 2000.

\bibitem{gadbury2004individual}
Gary~L Gadbury, Hari~K Iyer, and Jeffrey~M Albert.
\newblock Individual treatment effects in randomized trials with binary
  outcomes.
\newblock {\em Journal of Statistical Planning and Inference}, 121(2):163--174,
  2004.

\bibitem{gadbury2001evaluating}
Gary~L Gadbury, Hari~K Iyer, and David~B Allison.
\newblock Evaluating subject-treatment interaction when comparing two
  treatments.
\newblock {\em Journal of Biopharmaceutical Statistics}, 11(4):313--333, 2001.

\bibitem{holland1986statistics}
Paul~W Holland.
\newblock Statistics and causal inference.
\newblock {\em Journal of the American statistical Association},
  81(396):945--960, 1986.

\bibitem{imai2013estimating}
Kosuke Imai and Marc Ratkovic.
\newblock Estimating treatment effect heterogeneity in randomized program
  evaluation.
\newblock {\em The Annals of Applied Statistics}, 7(1):443--470, 2013.

\bibitem{poulson2012treatment}
Robert~S Poulson, Gary~L Gadbury, and David~B Allison.
\newblock Treatment heterogeneity and individual qualitative interaction.
\newblock {\em The American Statistician}, 66(1):16--24, 2012.

\bibitem{rosenbaum1983central}
Paul~R Rosenbaum and Donald~B Rubin.
\newblock The central role of the propensity score in observational studies for
  causal effects.
\newblock {\em Biometrika}, 70(1):41--55, 1983.

\bibitem{rubin1974estimating}
Donald~B Rubin.
\newblock Estimating causal effects of treatments in randomized and
  nonrandomized studies.
\newblock {\em Journal of Educational Psychology}, 66(5):688--701, 1974.

\bibitem{rubin1980randomization}
Donald~B Rubin.
\newblock Randomization analysis of experimental data: The fisher randomization
  test comment.
\newblock {\em Journal of the American Statistical Association},
  75(371):591--593, 1980.

\bibitem{shen2013treatment}
Changyu Shen, Jaesik Jeong, Xiaochun Li, Peng~Sheng Chen, and Alfred Buxton.
\newblock Treatment benefit and treatment harm rate to characterize
  heterogeneity in treatment effect.
\newblock {\em Biometrics}, 69(3):724--731, 2013.

\bibitem{doi:10.1056/NEJMoa1001288}
{\relax The ACCORD Study Group and ACCORD Eye Study Group}.
\newblock Effects of medical therapies on retinopathy progression in type 2
  diabetes.
\newblock {\em New England Journal of Medicine}, 363:233--244, 2010.

\bibitem{vickers1998homeopathic}
Andrew~J Vickers, Peter Fisher, Claire Smith, Sheena~E Wyllie, and Rebecca
  Rees.
\newblock Homeopathic arnica 30$\times$ is ineffective for muscle soreness
  after long-distance running: A randomized, double-blind, placebo-controlled
  trial.
\newblock {\em The Clinical Journal of Pain}, 14(3):227--231, 1998.

\bibitem{yin2016using}
Yunjian Yin and Xiao~Hua Zhou.
\newblock Using secondary outcome to sharpen inference in characterizing
  heterogeneity.
\newblock {\em Submitted for pulication}, 2016.

\bibitem{yin2016assessing}
Yunjian Yin, Xiao~Hua Zhou, Zhi Geng, and Fang Lu.
\newblock Assessing the heterogeneity of treatment effects by identifying the
  treatment benefit and treatment harm rate.
\newblock {\em Submitted for pulication}, 2016.

\bibitem{zhang2013assessing}
Zhiwei Zhang, Chenguang Wang, Lei Nie, and Guoxing Soon.
\newblock Assessing the heterogeneity of treatment effects via potential
  outcomes of individual patients.
\newblock {\em Journal of the Royal Statistical Society: Series C (Applied
  Statistics)}, 62(5):687--704, 2013.

\end{thebibliography}

\end{document}


\maketitle

\section{APPENDIX A: Proof of Formulas (F1), (F2), (F3) and (F4)}
\subsection{Proof of Formulas (F1) and (F2)}
\begin{eqnarray*}
    &&\mathrm{TBR}_{c}= P(Y_{1}-Y_{0}>c)  \\
   &=& P\big((\alpha_{1,0}-\alpha_{0,0})+(\alpha_{1,1}-\alpha_{0,1})^{T}X+(\alpha_{1,2}-\alpha_{0,2})U
   +(\alpha_{1,3}-\alpha_{0,3})^{T}XU+(\epsilon_{1}-\epsilon_{0}\big)>c) \\
   &=& P\big(\frac{\epsilon_{0}-\epsilon_{1}}{\sqrt{\sigma_{0}^{2}+\sigma_{1}^{2}}}<
   \frac{(\alpha_{1,0}-\alpha_{0,0})+(\alpha_{1,1}-\alpha_{0,1})^{T}X
   +(\alpha_{1,2}-\alpha_{0,2})U+(\alpha_{1,3}-\alpha_{0,3})^{T}XU-c}{\sqrt{\sigma_{0}^{2}+\sigma_{1}^{2}}}\big) \\
   &=&\int\int\Phi\Big(\frac{(\alpha_{1,0}-\alpha_{0,0})+(\alpha_{1,1}-\alpha_{0,1})^{T}x
   +(\alpha_{1,2}-\alpha_{0,2})u+(\alpha_{1,3}-\alpha_{0,3})^{T}xu-c}{\sqrt{\sigma_{0}^{2}+\sigma_{1}^{2}}}\Big)f_{U}(u)f_{X}(x)dudx\\
   &=&\int\int \Phi\Big(\big(w_{1}+w_{2}u\big)/w_{3}\Big)f_{U}(u)f_{X}(x)dudx \\
   &=&\int\int\int_{-\infty}^{(w_{1}+w_{2}u)/w_{3}}\frac{1}{\sqrt{2\pi}}\exp\big(-s^{2}/2\big)f_{U}(u)f_{X}(x)dsdudx\\
   &=&\int\int\int_{-\infty}^{0}\frac{1}{2\pi}\exp\Big[-\frac{1}{2w_{3}^{2}}\Big\{(w_{2}^{2}+w_{3}^{2})\big(u+\frac{w_{2}(w_{3}s+w_{1})}{w_{2}^{2}+w_{3}^{2}}\big)^{2}
   +\frac{w_{3}^{2}(w_{3}s+w_{1})^{2}}{w_{2}^{2}+w_{3}^{2}}\Big\}\Big]f_{X}(x)dsdudx\\
   &=&\int\int_{-\infty}^{0}\frac{1}{\sqrt{2\pi}}\sqrt{\frac{w_{3}^{2}}{w_{2}^{2}+w_{3}^{2}}}
   \exp\Big\{-\frac{(w_{3}s+w_{1})^{2}}{2(w_{2}^{2}+w_{3}^{2})}\Big\}f_{X}(x)dsdx\\
   &=&\int\Phi\Big(\frac{w_{1}}{\sqrt{w_{2}^{2}+w_{3}^{2}}}\Big)f_{X}(x)dx,
\end{eqnarray*}
where $f_{U}(\cdot)$ and $f_{X}(\cdot)$ are the density functions of $U$ and $X$ respectively, $w_{1}=(\alpha_{1,0}-\alpha_{0,0})+(\alpha_{1,1}-\alpha_{0,1})^{T}x-c,w_{2}=(\alpha_{1,2}-\alpha_{0,2})+(\alpha_{1,3}-\alpha_{0,3})^{T}x,
w_{3}=\sqrt{\sigma_{0}^{2}+\sigma_{1}^{2}}$.
Similarly, we can derive the form for $\mathrm{THR}_{c}$.

\subsection{Proof of Formulas (F3) and (F4)}
Let
$$K(\alpha_{t},x,u)=\alpha_{t,0}+\alpha_{t,1}^{T}x+\alpha_{t,2}u+\alpha_{t,3}^{T}xu.$$
Then
\begin{eqnarray*}
 &&\mathrm{TBR}  = \int\int \big\{1-\Phi\big(K(\alpha_{0},x,u)\big)\big\}\Phi\big(K(\alpha_{1},x,u)\big)f_{X}(x)f_{U}(u)dxdu \\
   &=& \int\int \Big\{\int^{\infty}_{K(\alpha_{0},x,u)}\frac{1}{\sqrt{2\pi}}\exp(-s_{0}^{2}/2)ds_{0}\Big\}
\Big\{\int_{-\infty}^{K(\alpha_{1},x,u)}\frac{1}{\sqrt{2\pi}}\exp(-s_{1}^{2}/2)ds_{1}\Big\}f_{X}(x)f_{U}(u)dxdu \\
   &=& \int\int\int_{-\infty}^{0}\int_{0}^{\infty}\frac{1}{(2\pi)^{3/2}}
   \exp\Big\{-\frac{(s_{0}+K(\alpha_{0},x,u))^{2}+(s_{1}+K(\alpha_{1},x,u))^{2}+u^{2}}{2}\Big\}f_{X}(x)ds_{0}ds_{1}dudx
\end{eqnarray*}
Let $K_{1}(\alpha_{t},x)=\alpha_{t,0}+\alpha_{t,1}^{T}x,K_{2}(\alpha_{t},x)=\alpha_{t,2}+\alpha_{t,3}^{T}x$, thus $K(\alpha_{t},x,u)=K_{1}(\alpha_{t},x)+u*K_{2}(\alpha_{t},x)$. Then the term in $\exp\big(-\frac{1}{2}(\cdot)\big)$ can be arranged as
\begin{eqnarray*}
   && \{s_{0}+K(\alpha_{0},x,u)\}^{2}+\{s_{1}+K(\alpha_{1},x,u)\}^{2}+u^{2} \\
   &=& \big\{1+K_{2}(\alpha_{0},x)^{2}+K_{2}(\alpha_{1},x)^{2}\big\}u^{2}
   +2\big\{(s_{0}+K_{1}(\alpha_{0},x))K_{2}(\alpha_{0},x)+(s_{1}+K_{1}(\alpha_{1},x))K_{2}(\alpha_{1},x)\big\}u \\
   &&+\{s_{0}+K_{1}(\alpha_{0},x)\}^{2}+\{s_{1}+K_{1}(\alpha_{1},x)\}^{2}  \\
   &=& \big\{1+K_{2}(\alpha_{0},x)^{2}+K_{2}(\alpha_{1},x)^{2}\big\}
   \big\{u+\frac{(s_{0}+K_{1}(\alpha_{0},x))K_{2}(\alpha_{0},x)+(s_{1}+K_{1}(\alpha_{1},x))K_{2}(\alpha_{1},x)}
{1+K_{2}(\alpha_{0},x)^{2}+K_{2}(\alpha_{1},x)^{2}}\big\}^{2}\\
    &&+\frac{1}{1+K_{2}(\alpha_{0},x)^{2}+K_{2}(\alpha_{1},x)^{2}}
    \Big[\{s_{0}+K_{1}(\alpha_{0},x)\}^{2}\{1+K_{2}(\alpha_{1},x)^{2}\}\\
    &&\hspace{5.7cm}+\{s_{1}+K_{1}(\alpha_{1},x)\}^{2}\{1+K_{2}(\alpha_{0},x)^{2}\}\\
    &&\hspace{5.7cm}-2\{s_{0}+K_{1}(\alpha_{0},x)\}K_{2}(\alpha_{0},x)\{s_{1}+K_{1}(\alpha_{1},x)\}K_{2}(\alpha_{1},x)\Big]
\end{eqnarray*}
So
$$\mathrm{TBR}=\int\int_{0}^{\infty}\int_{-\infty}^{0}\frac{1}{(2\pi)S}\exp\Big(-\frac{F}{2}\Big)f_{X}(x)ds_{0}ds_{1}dx,$$
where
$$S^{2}=1+K_{2}(\alpha_{0},x)^{2}+K_{2}(\alpha_{1},x)^{2},$$
\begin{eqnarray*}
 F &=& \Big[\{s_{0}+K_{1}(\alpha_{0},x)\}^{2}\{1+K_{2}(\alpha_{1},x)^{2}\}+\{s_{1}+K_{1}(\alpha_{1},x)\}^{2}\{1+K_{2}(\alpha_{0},x)^{2}\}\\
    &&-2\{s_{0}+K_{1}(\alpha_{0},x)\}K_{2}(\alpha_{0},x)\{s_{1}+K_{1}(\alpha_{1},x)\}K_{2}(\alpha_{1},x)\Big]/S^{2} \\
  &=&\Big\{(s_{0},s_{1})-\mu\Big\}\Sigma^{-1}\Big\{(s_{0},s_{1})-\mu\Big\}^{T}
\end{eqnarray*}
$$\mu=(-K_{1}(\alpha_{0},x),~-K_{1}(\alpha_{1},x)),$$
$$\Sigma=\left(
                \begin{array}{cc}
                  1+K_{2}(\alpha_{0},x)^{2}  & K_{2}(\alpha_{0},x)K_{2}(\alpha_{1},x) \\
                  K_{2}(\alpha_{0},x)K_{2}(\alpha_{1},x) & 1+K_{2}(\alpha_{1},x)^{2} \\
                \end{array}
              \right).$$
Thus,
$$\mathrm{TBR}=\int \Phi_{2}\big((0,\infty),(-\infty,0);\mu,\Sigma\big)f_{X}(x)dx,$$
where $\Phi_{2}(A_{0},A_{1};\mu,\Sigma)$ is the distribution function of bivariate normal vector with mean $\mu$, covariance matrix $\Sigma$ and integral region $A_{0}\times A_{1}$.
Similarly, we can derive the form for THR.

\section{APPENDIX B: Sufficient and necessary identification conditions of $(g_{t}(X),h_{t}(X))$}
\subsection{Continuous case}
$$\left\{
  \begin{array}{ll}
Y_{t}=g_{t}(X)+Uh_{t}(X)+\epsilon_{t},\\
\epsilon_{t}\bot(X,U),\epsilon_{t}\sim N(0,\sigma_{t}^{2}),U\sim N(0,1),~h(0)>0,
\end{array}
\right.$$
for $t=0,1$, where $X=(X_{1},\cdots,X_{p})^{T}$. Additional to Assumption 2, the following Condition A. is the sufficient and necessary condition to identify $(g_{0}(X),h_{0}(X),\sigma^{2}_{0},g_{1}(X),h_{1}(X),\sigma^{2}_{1})$.
\begin{itemize}
  \item [\textbf{A.:}] $h_{t}(X)$ belongs to the family $\mathcal{S}$ for $t=0,1$,
\end{itemize}
where
$$\mathcal{S}=\big\{h(X):\mathrm{h(X)~can~be~identified~if~h(X)h'(X)~is~known.}\big\}$$~\\
\textbf{Proof: } Without loss of generality, we can assume $U$ follows standard normal distribution.
Since $E[Y|X,T=t]=E[Y_{t}|X]=g_{t}(X)$, we can identify $g_{t}(X)$ and we have
$$\big(Y-g_{t}(X)\big)\big|\big(X,T=t\big)\sim N(0,h^{2}_{t}(X)+\sigma^{2}_{t}).$$
Thus $A_{t}(X)=h_{t}^{2}(X)+\sigma_{t}^{2}$ can also be identified, so is $A_{t}'(X)=h_{t}(X)h_{t}'(X)$.

Next we show that Condition A is sufficient and necessary to identify $h_{t}(x),t=0,1.$ It is easy to see that
if $h_{t}(X)$ belongs to $\mathcal{S}(X)$, then $h_{t}(X)$ is also identified. On the other hand,
if $h_{t}(X)$ does not belong to $\mathcal{S}(X)$, then $h_{t}(X)$ can not be decided uniquely from $h_{t}(X)h_{t}'(X)$. Besides, knowing $h_{t}(X)h_{t}'(X)$ is equivalent to knowing $h_{t}^{2}(X)$ up to a constant, i.e., $h_{t}^{2}(X_{1})-h_{t}^{2}(X_{2})$ for all $X_{1},X_{2}$. Note that $\Big(Y-g_{t}(X)\Big)\Big|(X,T=t)\sim N(0,h_{t}^{2}(X)+\sigma_{t}^{2}),$ the distribution of $Y-g_{t}(X)$ condition on $(X,T=t)$ is determined by the variance, so all the information we have about $h_{t}(X)$ is $h_{t}^{2}(X)+\sigma_{t}^{2}$, which is the same as knowing $h_{t}^{2}(X_{1})-h_{t}^{2}(X_{2})$ for all $X_{1},X_{2}$. Thus, we can not identify $h_{t}(X)$.

So the sufficient and necessary condition is that $h_{t}(X)\in\mathcal{S}(X)$ for $t=0,1$. \qed\\\\

\noindent\textbf{Lemma 1.} When $h(X)=h(X;\eta)=\eta_{0}+\eta_{1}^{T}X$, where $\eta=(\eta_{0},\eta_{1}^{T})^{T},\eta_{1}=\big(\eta_{1,1},\cdots,\eta_{1,p}\big)^{T},$ we can have $h(X)\in\mathcal{S}$.\\

\noindent\textbf{Proof. }Since
$$h(X)h'(X)=\big(\eta_{0}+\eta_{1}^{T}X\big)\eta_{1}=\eta_{0}\eta_{1}+\eta_{1}\eta_{1}^{T}X.$$
So $(\eta_{0}\eta_{1},\eta_{1}\eta_{1}^{T})$ can be identified if $h(X)h'(X)$ is known.
Besides, $h(0)=\eta_{0}>0$, so the sign of every component of $\eta_{1}$ can be determined since we know $\eta_{0}\eta_{1}.$ Then $\eta_{1}$ can be identified since we know the diagonal elements of $\eta_{1}\eta_{1}^{T}.$ Then $\eta_{0}$ can also be identified from $\eta_{0}\eta_{1}$. Thus $(\eta_{0},\eta_{1})$ is identifiable, so is $h(X)$.
This completes the proof of the continuous part in Theorem 1. \qed

\subsection{Binary case}
$$\left\{
  \begin{array}{ll}
Y_{t}^{*}=g_{t}(X)+Uh_{t}(X)+\epsilon_{t},\\
Y_{t}=I(Y_{t}^{*}>0),\\
\epsilon_{t}\bot(X,U),\epsilon_{t}\sim N(0,\sigma^{2}),U\sim N(\mu_{U},\sigma^{2}_{U}),
\end{array}
\right.$$
where $t=0,1$. Additional to Assumption 2, the following Condition B. is the sufficient and necessary condition to identify $(g_{0}(X),h_{0}(X),g_{1}(X),h_{1}(X))$.
\begin{itemize}
  \item [\textbf{B.:}] $(g_{t}(X),h_{t}(X))$ belongs to the family $\big(\mathcal{S}_{1}(X),\mathcal{S}_{2}(X)\big)$ for $t=0,1$,
\end{itemize}
 where
\begin{eqnarray*}
   && \big(\mathcal{S}_{1}(X),\mathcal{S}_{2}(X)\big) \\
   &=&  \Big\{\big(g(X;\alpha_{1}),h(X;\alpha_{2})\big)\big|(\alpha_{1},\alpha_{2})\in\mathcal{A},
\forall (\alpha_{1},\alpha_{2})\neq(\beta_{1},\beta_{2})\in\mathcal{A},
\frac{g(X;\alpha_{1})}{\sqrt{1+h^{2}(X;\alpha_{2})}}\neq\frac{g(X;\beta_{1})}{\sqrt{1+h^{2}(X;\beta_{2})}}\Big\}
\end{eqnarray*}
\textbf{Proof: }Without loss of generality, we can assume U follows standard normal distribution. Since $P(Y=1|X,U,T=t)=\Phi\big(g_{t}(X)+Uh_{t}(X)\big),$ we have
$$P(Y=1|X,T=t)=\Phi\Big(\frac{g_{t}(X)}{\sqrt{1+h_{t}^{2}(X)}}\Big),$$
It is easy to see that $(g_{0}(X),h_{0}(X),g_{1}(X),h_{1}(X))$ can be identified if and only if the Condition B holds. \qed\\\\

\noindent\textbf{Lemma 2.} When $g(X)=g(X;\alpha)=\alpha_{0}+\alpha_{1}^{T}X,~h(X)=h(X;\alpha)=\alpha_{2}+\alpha_{3}^{T}X$ with $(\alpha_{0},\alpha_{1})\neq0, \alpha_{2}>0, \alpha_{3}\neq0,$ where $\alpha=(\alpha_{0},\alpha_{1}^{T},\alpha_{2},\alpha_{3}^{T})^{T}, \alpha_{1}=(\alpha_{1,1},\cdots,\alpha_{1,p})^{T},~\alpha_{3}=(\alpha_{3,1},\cdots,\alpha_{3,p})^{T}$, if the following Condition C. holds,
we can have $\{g(X),h(X)\}\in\{\mathcal{S}_{1}(X),\mathcal{S}_{2}(X)\}$.
\begin{itemize}
  \item [\textbf{C.:}] $\mathrm{There ~exists~ linear~independent}~(\tau_{1},\cdots,\tau_{p})\subset\mathcal{X}, \mathrm{where} ~\mathcal{X}~ \mathrm{is~the~domain~of}~ X,~ s.~t.~ P(Y=1|X=\tau_{i})=0,i=1,..,p.$
\end{itemize}

\noindent\textbf{Proof.} It is enough to show that if $\alpha=(\alpha_{0},\alpha_{1}^{T},\alpha_{2},\alpha_{3}^{T})^{T},\beta=(\beta_{0},\beta_{1}^{T},\beta_{2},\beta_{3}^{T})^{T}$ satisfy:
\begin{equation}\label{P1}
\frac{\alpha_{0}+\alpha_{1}^{T}X}{\sqrt{1+(\alpha_{2}+\alpha_{3}^{T}X)^{2}}}=\frac{\beta_{0}+\beta_{1}^{T}X}{\sqrt{1+(\beta_{2}+\beta_{3}^{T}X)^{2}}}, ~\forall X\in\mathcal{X},
\end{equation}
then $\alpha=\beta$. To keep the same signs on both sides, the following two subsets of a hyperplane $(H_{0},H_{1})$ must be the same,
$$H_{0}=\{X\subset\mathcal{X}|\alpha_{0}+\alpha_{1}^{T}X=0\},~H_{1}=\{X\subset\mathcal{X}|\beta_{0}+\beta_{1}^{T}X=0\},$$
since there exists linear independent $(\tau_{1},\cdots,\tau_{p})\subset\mathcal{X}$ such that $P(Y=1|X=\tau_{i})=0,i=1,..,p$, thus, the following two hyperplane $(\widetilde{H}_{0},\widetilde{H}_{1})$ must be the same,
$$\widetilde{H}_{0}=\{X\subset\mathbb{R}^{p}|\alpha_{0}+\alpha_{1}^{T}X=0\},
~\widetilde{H}_{1}=\{X\subset\mathbb{R}^{p}|\beta_{0}+\beta_{1}^{T}X=0\},$$
which means $(\alpha_{0},\alpha_{1}^{T})=k(\beta_{0},\beta_{1}^{T})$, and $k\geq0$ since the signs on the two sides of equations (\ref{P1}) must be the same. And $(\alpha_{0},\alpha_{1})\neq0$ exclude the case $k=0$. Thus from equation (\ref{P1}) we have
$$k^{2}=\frac{1+(\alpha_{2}+\alpha_{3}^{T}X)^{2}}{1+(\beta_{2}+\beta_{3}^{T}X)^{2}}.$$
By arranging the equation above we can have
$$X^{T}(\alpha_{3}\alpha_{3}^{T}-k^{2}\beta_{3}\beta_{3}^{T})X+2(\alpha_{2}\alpha_{3}^{T}-k^{2}\beta_{2}\beta_{3}^{T})X+1+\alpha_{2}^{2}-k-k\beta_{2}^{2}=0.$$
So
\begin{subequations}
\begin{align}
\alpha_{3}\alpha_{3}^{T}-k^{2}\beta_{3}\beta_{3}^{T} &=0, \label{eq:parent1}\\
\alpha_{2}\alpha_{3}^{T}-k^{2}\beta_{2}\beta_{3}^{T} &=0,\label{eq:parent2} \\
1+\alpha_{2}^{2}-k^{2}-k^{2}\beta_{2}^{2}&=0. \label{eq:parent3}
\end{align}
Note that the two sides of equation (\ref{eq:parent1}) are both matrixes and the two sides of equation (\ref{eq:parent2}) are both vectors. Take the $(i,i)$ element of (\ref{eq:parent1}) and the i-th component of (\ref{eq:parent2}), with a little arrangement we can have
\begin{align}
\alpha_{3i}^{2}&=k^{2}\beta_{3i}^{2} \label{eq:sub1}\\
\alpha_{2}\alpha_{3i}&=k^{2}\beta_{2}\beta_{3i}\label{eq:sub2}\\
\alpha_{2}^{2}&=k^{2}+k^{2}\beta_{2}^{2}-1\label{eq:sub3}
\end{align}
\end{subequations}
$(\ref{eq:sub1})\cdot(\ref{eq:sub3})-(\ref{eq:sub2})^{2}=k^{2}\beta_{3i}^{2}(k^{2}-1)=0$, since $k>0$ we can have $k=1.$ And since $\alpha_{2},\beta_{2}\geq0,$ from (\ref{eq:parent3}) we have $\alpha_{2}=\beta_{2},$ then from (\ref{eq:parent2}) we have $\alpha_{3}=\beta_{3}$. Thus, $\alpha=\beta.$ This completes the proof of the binary part in Theorem 1. \qed

\section{APPENDIX C: Proof of Theorem 2}
The estimator $(\widehat{\theta}=(\widehat{\alpha}_{0,0},\widehat{\alpha}_{0,1}^{T},\widehat{\alpha}_{0,2},\widehat{\alpha}_{0,3}^{T},\widehat{\sigma}_{0}^{2},
\widehat{\alpha}_{1,0},\widehat{\alpha}_{1,1}^{T},\widehat{\alpha}_{1,2},\widehat{\alpha}_{1,3}^{T},\widehat{\sigma}_{1}^{2})^{T}$ maximize the following likelihood
\begin{eqnarray*}
  &&\ell=\log L(Y|X)\\ &=& \sum_{i=1}^{n}\sum_{t=0,1}\frac{1}{2}\bigg[I(T_{i}=t)\Big\{-\log(2\pi)-\log\big((\alpha_{t,2}+\alpha_{t,3}^{T}X_{i})^{2}+\sigma_{t}^{2}\big)
-\frac{(Y_{i}-\alpha_{t,0}-\alpha_{t,1}^{T}X_{i})^{2}}{(\alpha_{t,2}+\alpha_{t,3}^{T}X_{i})^{2}+\sigma_{t}^{2}}\Big\} \bigg]
\end{eqnarray*}
According to the M-estimator property, we can have
$$\widehat{\theta}-\theta=-\Big[P_{0}\big\{\frac{\partial^{2}}{\partial\theta\partial\theta^{T}}\psi(\theta,T,X,Y)\big\}\Big]^{-1}
P_{n}\big\{\frac{\partial}{\partial\theta}\psi(\theta,T,X,Y)\big\}+o_{p}(1/\sqrt{n}),$$
where $P_{0}$ and $P_{n}$ are the true mean and empirical mean respectively,
\begin{eqnarray*}
 &&\psi(T,X,Y;\theta)\\  &=& \sum_{t=0,1}\frac{1}{2}\bigg[I(T=t)\Big\{-\log(2\pi)-\log\big((\alpha_{t,2}+\alpha_{t,3}^{T}X)^{2}+\sigma_{t}^{2}\big)
-\frac{(Y-\alpha_{t,0}-\alpha_{t,1}^{T}X)^{2}}{(\alpha_{t,2}+\alpha_{t,3}^{T}X)^{2}+\sigma_{t}^{2}}\Big\} \bigg]
\end{eqnarray*}
Denote
$$m_{B}(X;\theta)=\Phi\Big(\frac{(\alpha_{1,0}-\alpha_{0,0})+(\alpha_{1,1}-\alpha_{0,1})^{T}X-c}
{\sqrt{((\alpha_{1,2}-\alpha_{0,2})+(\alpha_{1,3}-\alpha_{0,3})^{T}X)^{2}+(\sigma_{0}^{2}+\sigma_{1}^{2})}}\Big),$$
Then
\begin{eqnarray*}
 &&\sqrt{n}(\widehat{\mathrm{TBR}}_{c}-\mathrm{TBR}_{c})  = \sqrt{n}\big[P_{n}\{m_{B}(X;\widehat{\theta})\}-P_{0}\{m_{B}(X;\theta)\}\big] \\
   &=&  \sqrt{n}\big(P_{n}-P_{0}\big)\big\{m_{B}(X;\widehat{\theta})-m_{B}(X;\theta)\big\}
   +\sqrt{n}P_{0}\big\{m_{B}(X;\widehat{\theta})-m_{B}(X;\theta)\big\}\\
   &&+\sqrt{n}\big(P_{n}-P_{0}\big)\{m_{B}(X;\theta)\}\\
 &\equiv&V_{1}+V_{2}+V_{3}
\end{eqnarray*}
The first term $V_{1}$ is a negligible second order term, which is $o_{p}(1)$.
For the second item $V_{2}$, we use the functional delta method,
\begin{eqnarray*}
   &&\sqrt{n}P_{0}\big\{m_{B}(X;\widehat{\theta})-m_{B}(X;\theta)\big\}  \\
   &=& \sqrt{n}\Big[\frac{\partial}{\partial \theta}P_{0}\big\{m_{B}(X;\theta)\big\}\Big]^{T}\big(\widehat{\theta}-\theta\big)+o_{p}(1) \\
   &=& -\sqrt{n}\Big[\frac{\partial}{\partial \theta}P_{0}\big\{m_{B}(X;\theta)\big\}\Big] ^{T} \Big[P_{0}\big\{\frac{\partial^{2}}{\partial\theta\partial\theta^{T}}\psi(T,X,Y;\theta)\big\}\Big]^{-1}
\Big[P_{n}\big\{\frac{\partial}{\partial\theta}\psi(T,X,Y;\theta)\big\}\Big]+o_{p}(1)\\
   &=&- \sqrt{n}\Big[\frac{\partial}{\partial \theta}P_{0}\big\{m_{B}(X;\theta)\big\}\Big] ^{T} \Big[P_{0}\big\{\frac{\partial^{2}}{\partial\theta\partial\theta^{T}}\psi(T,X,Y;\theta)\big\}\Big]^{-1}
\Big[\big[P_{n}-P_{0}\big]\big\{\frac{\partial}{\partial\theta}\psi(T,X,Y;\theta)\big\}\Big]+o_{p}(1)\\
   &=&- \sqrt{n}\bigg[P_{n}-P_{0}\bigg]\bigg[\Big[\frac{\partial}{\partial \theta}P_{0}\big\{m_{B}(X;\theta)\big\}\Big] ^{T} \Big[P_{0}\big\{\frac{\partial^{2}}{\partial\theta\partial\theta^{T}}\psi(T,X,Y;\theta)\big\}\Big]^{-1}
\frac{\partial}{\partial\theta}\psi(T,X,Y;\theta)\bigg]+o_{p}(1)
\end{eqnarray*}
The penultimate equality holds because $P_{0}\big[\frac{\partial}{\partial\theta}\psi(T,X,Y;\theta)\big]=0$ (The derivative of the log-likelihood is zero at the true value of the parameter). Thus, we can have
\begin{eqnarray*}
   && \sqrt{n}\big(\widehat{\mathrm{TBR}}_{c}-\mathrm{TBR}_{c}\big) \\
   &=& \sqrt{n}\bigg[P_{n}-P_{0}\bigg]\bigg[-\Big[\frac{\partial}{\partial \theta}P_{0}\big\{m_{B}(X;\theta)\big\}\Big] ^{T} \Big[P_{0}\big\{\frac{\partial^{2}}{\partial\theta\partial\theta^{T}}\psi(T,X,Y;\theta)\big\}\Big]^{-1}
\frac{\partial}{\partial\theta}\psi(T,X,Y;\theta)+m_{B}(X;\theta)\bigg]\\&&+o_{p}(1)
\end{eqnarray*}

Thus, we can have
$$\sqrt{n}\big(\widehat{\mathrm{TBR}}_{c}-\mathrm{TBR}_{c}\big)\stackrel{d}{\longrightarrow} N(0,\sigma_{cB}^{2}(\theta)),$$
where
$$\sigma_{cB}^{2}(\theta)=\mathrm{var}\bigg[-\Big[P_{0}\big\{\frac{\partial}{\partial \theta}m_{B}(X;\theta)\big\}\Big] ^{T} \Big[P_{0}\big\{\frac{\partial^{2}}{\partial\theta\partial\theta^{T}}\psi(T,X,Y;\theta)\big\}\Big]^{-1}
\frac{\partial}{\partial\theta}\psi(T,X,Y;\theta)+m_{B}(X;\theta)\bigg].$$
Similarly, we can have
$$\sqrt{n}\big(\widehat{\mathrm{THR}}_{c}-\mathrm{THR}_{c}\big)\stackrel{d}{\longrightarrow} N(0,\sigma_{cH}^{2}(\theta)),$$
where
$$\sigma_{cH}^{2}(\theta)=\mathrm{var}\bigg[-\Big[P_{0}\big\{\frac{\partial}{\partial \theta}m_{H}(X;\theta)\big)\Big] ^{T} \Big[P_{0}\big\{\frac{\partial^{2}}{\partial\theta\partial\theta^{T}}\psi(T,X,Y;\theta)\big\}\Big]^{-1}
\frac{\partial}{\partial\theta}\psi(T,X,Y;\theta)+m_{H}(X;\theta)\bigg].$$
$$m_{H}(X;\theta)=\Phi\Big(\frac{(\alpha_{0,0}-\alpha_{1,0})+(\alpha_{0,1}-\alpha_{1,1})^{T}X-c}
{\sqrt{((\alpha_{0,2}-\alpha_{1,2})+(\alpha_{0,3}-\alpha_{1,3})^{T}X)^{2}+(\sigma_{0}^{2}+\sigma_{1}^{2})}}\Big).$$
$\sigma_{cB}^{2}(\theta)$ and $\sigma_{cH}^{2}(\theta)$ can be estimated by
$$\widehat{\sigma}_{cB}^{2}(\widehat{\theta})=\widehat{\mathrm{var}}\bigg[\Big[-P_{n}\big\{\frac{\partial}{\partial \theta}m_{B}(X;\widehat{\theta})\big\}\Big]^{T}\Big[P_{n}\big\{\frac{\partial^{2}}{\partial\theta\partial\theta^{T}}\psi(T,X,Y;\widehat{\theta})\big\}\Big]^{-1}
\frac{\partial}{\partial\theta}\psi(T,X,Y;\widehat{\theta})+m_{B}(X;\widehat{\theta})\bigg],$$
$$\widehat{\sigma}_{cH}^{2}(\widehat{\theta})=\widehat{\mathrm{var}}\bigg[\Big[-P_{n}\big\{\frac{\partial}{\partial \theta}m_{H}(X;\widehat{\theta})\big\}\Big]^{T}\Big[P_{n}\big\{\frac{\partial^{2}}{\partial\theta\partial\theta^{T}}\psi(T,X,Y;\widehat{\theta})\big\}\Big]^{-1}
\frac{\partial}{\partial\theta}\psi(T,X,Y;\widehat{\theta})+m_{H}(X;\widehat{\theta})\bigg],$$
where
$$\widehat{\mathrm{var}}[O]=P_{n}\big\{O-P_{n}(O)\big\}^{2}.$$
\section{APPENDIX D: Proof of Theorem 3}
The estimator $\widehat{\theta}=(\widehat{\alpha}_{0,0},\widehat{\alpha}_{0,1}^{T},\widehat{\alpha}_{0,2},\widehat{\alpha}_{0,3}^{T},
\widehat{\alpha}_{1,0},\widehat{\alpha}_{1,1}^{T},\widehat{\alpha}_{1,2},\widehat{\alpha}_{1,3}^{T})^{T}$ maximize the following likelihood
\begin{eqnarray*}
  &&\ell=\log L(Y|X)
   = \sum_{i=1}^{n}\sum_{t=0,1} \bigg[I(T_{i}=t)\Big\{Y_{i}\log\big(G(X_{i};\theta_{t})\big)+(1-Y_{i})\log\big(1-G(X_{i};\theta_{t})\big)\Big\} \bigg],
\end{eqnarray*}
where
$$G(X;\theta_{t})=\Phi\Big(\frac{\alpha_{t,0}+\alpha_{t,1}^{T}X}{\sqrt{1+(\alpha_{t,2}+\alpha_{t,3}^{T}X)^{2}}}\Big).$$
According to the M-estimator property, we can have
$$\widehat{\theta}-\theta=-\Big[P_{0}\big\{\frac{\partial^{2}}{\partial\theta\partial\theta^{T}}\psi(T,X,Y;\theta)\big\}\Big]^{-1}
\frac{1}{n}\sum_{i=1}^{n}\frac{\partial}{\partial\theta}\psi(T_{i},X_{i},Y_{i};\theta)+o_{p}(1/\sqrt{n}),$$
where
\begin{eqnarray*}
  &&\psi(T,X,Y;\theta)=\sum_{t=0,1}\bigg[I(T=t)\Big\{Y\log\big(G(X;\theta_{t})\big)+(1-Y)\log\big(1-G(X;\theta_{t})\big)\Big\}\bigg].
\end{eqnarray*}

Denote
$$g_{01}(X;\theta)=\Phi_{2}\big((0,\infty),(-\infty,0);\mu(X;\theta),\Sigma(X;\theta)\big),$$
where
$$\mu(X;\theta)=\big(\mu_{0}(X;\theta),\mu_{1}(X;\theta)\big)=\big(-\alpha_{0,0}-\alpha_{0,1}^{T}X,~-\alpha_{1,0}-\alpha_{1,1}^{T}X\big),$$
$$\Sigma(X;\theta)=\left(
                \begin{array}{cc}
                  1+(\alpha_{0,2}+\alpha_{0,3}^{T}X)^{2}  & (\alpha_{0,2}+\alpha_{0,3}^{T}X)(\alpha_{1,2}+\alpha_{1,3}^{T}X) \\
                  (\alpha_{0,2}+\alpha_{0,3}^{T}X)(\alpha_{1,2}+\alpha_{1,3}^{T}X) & 1+(\alpha_{1,2}+\alpha_{1,3}^{T}X)^{2} \\
                \end{array}
              \right).$$
and $\Phi_{2}(A_{0},A_{1};\mu,\Sigma)$ is the distribution function of bivariate normal vector with mean $\mu$, covariance matrix $\Sigma$ in the integration region $A_{0}\times A_{1}$, i.e.,
$$\Phi_{2}(A_{0},A_{1};\mu,\Sigma)=\int\int_{A_{0}\times A_{1}}\frac{1}{2\pi|\Sigma|^{1/2}}\exp\big\{-\frac{1}{2}(s_{0}-\mu_{0},s_{1}-\mu_{1})\Sigma^{-1}(s_{0}-\mu_{0},s_{1}-\mu_{1})^{T}\big\}ds_{0}ds_{1}.$$
Then we have
\begin{eqnarray*}
 &&\sqrt{n}(\widehat{\mathrm{TBR}}-\mathrm{TBR})  = \sqrt{n}\big[P_{n}\{g_{01}(X;\widehat{\theta})\}-P_{0}\{g_{01}(X;\theta)\}\big] \\
   &=&  \sqrt{n}\big(P_{n}-P_{0}\big)\big(g_{01}(X;\widehat{\theta})-g_{01}(X;\theta)\big)
   +\sqrt{n}P_{0}\big\{g_{01}(X;\widehat{\theta})-g_{01}(X;\theta)\big\}+\sqrt{n}\big(P_{n}-P_{0}\big)\{g_{01}(X;\theta)\}\\
 &=&V_{1}+V_{2}+V_{3}
\end{eqnarray*}
The first term $V_{1}$ is a negligible second order term, which is $o_{p}(1)$.
For the second item $V_{2}$, we use the functional delta method,
\begin{eqnarray*}
   &&\sqrt{n}P_{0}\big\{g_{01}(X;\widehat{\theta})-g_{01}(X;\theta)\big\}  \\
   &=& \sqrt{n}\Big[\frac{\partial}{\partial \theta}P_{0}\big\{g_{01}(X;\theta)\big\}\Big]^{T}\big(\widehat{\theta}-\theta\big)+o_{p}(1) \\
   &=& -\sqrt{n}\Big[\frac{\partial}{\partial \theta}P_{0}\big\{g_{01}(X;\theta)\big\}\Big] ^{T} \Big[P_{0}\big\{\frac{\partial^{2}}{\partial\theta\partial\theta^{T}}\psi(T,X,Y;\theta)\big\}\Big]^{-1}
\Big[P_{n}\big\{\frac{\partial}{\partial\theta}\psi(T,X,Y;\theta)\big\}\Big]+o_{p}(1)\\
   &=&- \sqrt{n}\Big[\frac{\partial}{\partial \theta}P_{0}\big\{g_{01}(X;\theta)\big\}\Big] ^{T} \Big[P_{0}\big\{\frac{\partial^{2}}{\partial\theta\partial\theta^{T}}\psi(T,X,Y;\theta)\big\}\Big]^{-1}
\Big[\big[P_{n}-P_{0}\big]\big\{\frac{\partial}{\partial\theta}\psi(T,X,Y;\theta)\big\}\Big]+o_{p}(1)\\
   &=&- \sqrt{n}\bigg[P_{n}-P_{0}\bigg]\bigg[\Big[\frac{\partial}{\partial \theta}P_{0}\big\{g_{01}(X;\theta)\big\}\Big] ^{T} \Big[P_{0}\big\{\frac{\partial^{2}}{\partial\theta\partial\theta^{T}}\psi(T,X,Y;\theta)\big\}\Big]^{-1}
\frac{\partial}{\partial\theta}\psi(T,X,Y;\theta)\bigg]+o_{p}(1)
\end{eqnarray*}
The penultimate equality holds because the derivative of the log-likelihood is zero at the true value of the parameter, i.e., $P_{0}\big[\frac{\partial}{\partial\theta}\psi(T,X,Y;\theta)\big]=0$. Thus, we can have
\begin{eqnarray*}
   && \sqrt{n}\big(\widehat{\mathrm{TBR}}-\mathrm{TBR}\big) \\
   &=& \sqrt{n}\bigg[P_{n}-P_{0}\bigg]\bigg[-\Big[\frac{\partial}{\partial \theta}P_{0}\big\{g_{01}(X;\theta)\big\}\Big] ^{T} \Big[P_{0}\big\{\frac{\partial^{2}}{\partial\theta\partial\theta^{T}}\psi(T,X,Y;\theta)\big\}\Big]^{-1}
\frac{\partial}{\partial\theta}\psi(T,X,Y;\theta)+g_{01}(X;\theta)\bigg]\\&&+o_{p}(1)
\end{eqnarray*}
Thus, we can have
$$\sqrt{n}\big(\widehat{\mathrm{TBR}}-\mathrm{TBR}\big)\stackrel{d}{\longrightarrow} N(0,\sigma_{bB}^{2}(\theta)),$$
where
$$\sigma_{cB}^{2}(\theta)=\mathrm{var}\bigg[-\Big[\frac{\partial}{\partial \theta}P_{0}\big\{g_{01}(X;\theta)\big\}\Big]^{T} \Big[P_{0}\big\{\frac{\partial^{2}}{\partial\theta\partial\theta^{T}}\psi(T,X,Y;\theta)\big\}\Big]^{-1}
\frac{\partial}{\partial\theta}\psi(T,X,Y;\theta)+g_{01}(X;\theta)\bigg].$$
Similarly, we can have
$$\sqrt{n}\big(\widehat{\mathrm{THR}}-\mathrm{THR}\big)\stackrel{d}{\longrightarrow} N(0,\sigma_{bH}^{2}(\theta)),$$
where
$$\sigma_{bH}^{2}(\theta)=\mathrm{var}\bigg[-\Big[\frac{\partial}{\partial \theta}P_{0}\big\{g_{10}(X;\theta)\big\}\Big]^{T} \Big[P_{0}\big\{\frac{\partial^{2}}{\partial\theta\partial\theta^{T}}\psi(T,X,Y;\theta)\big\}\Big]^{-1}
\frac{\partial}{\partial\theta}\psi(T,X,Y;\theta)+g_{10}(X;\theta)\bigg].$$
$$g_{10}(X;\theta)=\Phi_{2}\big((-\infty,0),(0,\infty);\mu_{\mathrm{HTE}}(\theta,X),\Sigma_{\mathrm{HTE}}(\theta,X)\big),$$

$\sigma_{bB}^{2}(\theta)$ and $\sigma_{bH}^{2}(\theta)$ can be estimated by
$$\widehat{\sigma}_{cB}^{2}(\widehat{\theta})=\widehat{\mathrm{var}}\bigg[-\Big[\frac{\partial}{\partial \theta}P_{n}\big\{g_{01}(X;\widehat{\theta})\big\}\Big] ^{T} \Big[P_{n}\big\{\frac{\partial^{2}}{\partial\theta\partial\theta^{T}}\psi(T,X,Y;\widehat{\theta})\big\}\Big]^{-1}
\frac{\partial}{\partial\theta}\psi(T,X,Y;\widehat{\theta})+g_{01}(X;\widehat{\theta})\bigg],$$
$$\widehat{\sigma}_{cH}^{2}(\widehat{\theta})=\widehat{\mathrm{var}}\bigg[-\Big[\frac{\partial}{\partial \theta}P_{n}\big\{g_{10}(X;\widehat{\theta})\big\}\Big] ^{T} \Big[P_{n}\big\{\frac{\partial^{2}}{\partial\theta\partial\theta^{T}}\psi(T,X,Y;\widehat{\theta})\big\}\Big]^{-1}
\frac{\partial}{\partial\theta}\psi(T,X,Y;\widehat{\theta})+g_{10}(X;\widehat{\theta})\bigg],$$

\section{APPENDIX E: Estimation and Asymptotic Properties for The General Models}
All the results in linear models can be extended to general models, and they have the same expressions by replacing
$$g_{t}(x)=\alpha_{t,0}+\alpha_{t,1}x,~~~h_{t}(x)=\alpha_{t,2}+\alpha_{t,3}x.$$

The corresponding formulas of (F1),(F2),(F3) and (F4) for the general models are:
$$\left\{
    \begin{array}{ll}
\mathrm{TBR}_{c}&=\int\Phi\Big(\frac{\big(g_{1}(x)-g_{0}(x)\big)-c}
{\sqrt{\big(h_{1}(x)-h_{0}(x)\big)^{2}+\sigma_{0}^{2}+\sigma_{1}^{2}}}\Big)f_{X}(x)dx,\\
\mathrm{TBR}_{c}&=\int\Phi\Big(\frac{\big(g_{0}(x)-g_{1}(x)\big)-c}
{\sqrt{\big(h_{0}(x)-h_{1}(x)\big)^{2}+\sigma_{0}^{2}+\sigma_{1}^{2}}}\Big)f_{X}(x)dx,
    \end{array}
  \right.
$$
and
$$\left\{
  \begin{array}{ll}
\mathrm{TBR}&=\int \Phi_{b}\big(\widetilde{\mu}(x),\widetilde{\Sigma}(x)\big)f_{X}(x)dx,\\
\mathrm{THR}&=\int \Phi_{h}\big(\widetilde{\mu}(x),\widetilde{\Sigma}(x)\big)f_{X}(x)dx,
  \end{array}
\right.$$
where
$$\widetilde{\mu}(x)=-\big(g_{0}(x),~g_{1}(x)\big),$$
$$\widetilde{\Sigma}(x)=\left(
                \begin{array}{cc}
                  1+h_{0}^{2}(x)  & h_{0}(x)h_{1}(x) \\
                  h_{0}(x)h_{1}(x) & 1+h^{2}_{1}(x) \\
                \end{array}
              \right).$$

In estimation, we first build models for $g_{t}(X)$ and $h_{t}(X)$, respectively, denote as $g_{t}(X;\alpha_{t,1})$ and $h_{t}(X;\alpha_{t,2})$. Also denote $\psi(T,X,Y;\theta)$ as the log-density function, where $\theta=(\alpha_{0,1},\alpha_{0,2},\sigma^{2}_{0},\alpha_{1,1},\alpha_{1,2},\sigma^{2}_{1})^{T}$ in the continuous case and $\theta=(\alpha_{0,1},\alpha_{0,2},\alpha_{1,1},\alpha_{1,2})^{T}$ in the binary case. The estimation for $\theta$ can be obtained by maximizing $P_{n}[\psi(T,X,Y;\theta)]$, denote as $\widehat{\theta}$. Then $\mathrm{TBR},\mathrm{THR},\mathrm{TBR}_{c},\mathrm{THR}_{c}$ can be estimated by:
$$\left\{
    \begin{array}{ll}
\mathrm{TBR}_{c}&=P_{n}\Big[\Phi\Big(\frac{\big(g_{1}(X;\widehat{\alpha}_{1,1})-g_{0}(X;\widehat{\alpha}_{0,1})\big)-c}
{\sqrt{\big(h_{1}(X;\widehat{\alpha}_{1,2})-h_{0}(X;\widehat{\alpha}_{0,2})\big)^{2}+\widehat{\sigma}_{0}^{2}+\widehat{\sigma}_{1}^{2}}}\Big)\Big],\\
\mathrm{THR}_{c}&=P_{n}\Big[\Phi\Big(\frac{\big(g_{1}(X;\widehat{\alpha}_{0,1})-g_{0}(X;\widehat{\alpha}_{1,1})\big)-c}
{\sqrt{\big(h_{1}(X;\widehat{\alpha}_{0,2})-h_{0}(X;\widehat{\alpha}_{1,2})\big)^{2}+\widehat{\sigma}_{0}^{2}+\widehat{\sigma}_{1}^{2}}}\Big)\Big],\\
\mathrm{TBR}&=P_{n}\Big[\Phi_{b}\big(\widetilde{\mu}(X;\widehat{\theta}),\widetilde{\Sigma}(X;\widehat{\theta})\big)\Big],\\
\mathrm{THR}&=P_{n}\Big[\Phi_{h}\big(\widetilde{\mu}(X;\widehat{\theta}),\widetilde{\Sigma}(X;\widehat{\theta})\big)\Big],
  \end{array}
\right.$$
where
$$\widetilde{\mu}(X;\widehat{\theta})=\big(-g_{0}(X;\widehat{\alpha}_{0,1}),~-g_{1}(X;\widehat{\alpha}_{1,1})\big),$$
$$\widetilde{\Sigma}(X;\widehat{\theta})=\left(
                \begin{array}{cc}
                  1+h_{0}^{2}(X;\widehat{\alpha}_{0,2})  & h_{0}(X;\widehat{\alpha}_{0,2})h_{1}(X;\widehat{\alpha}_{1,2}) \\
                  h_{0}(X;\widehat{\alpha}_{0,2})h_{1}(X;\widehat{\alpha}_{1,2}) & 1+h^{2}_{1}(X;\widehat{\alpha}_{1,2}) \\
                \end{array}
              \right).$$~\\\\

For the continuous case, the asymptotic distribution is
\begin{eqnarray*}
   && \sqrt{n}\big(\widehat{\mathrm{TBR}}_{c}-\mathrm{TBR}_{c}\big) \\
   &=& \sqrt{n}\bigg[P_{n}-P_{0}\bigg]\bigg[-\Big[\frac{\partial}{\partial \theta}P_{0}\big\{\widetilde{m}_{B}(X;\theta)\big\}\Big]^{T} \Big[P_{0}\big\{\frac{\partial^{2}}{\partial\theta\partial\theta^{T}}\psi(T,X,Y;\theta)\big\}\Big]^{-1}
\frac{\partial}{\partial\alpha}\psi(T,X,Y;\theta)+\widetilde{m}_{B}(X;\theta)\bigg]\\&&+o_{p}(1),
\end{eqnarray*}
\begin{eqnarray*}
   && \sqrt{n}\big(\widehat{\mathrm{THR}}_{c}-\mathrm{THR}_{c}\big) \\
   &=& \sqrt{n}\bigg[P_{n}-P_{0}\bigg]\bigg[-\Big[\frac{\partial}{\partial \theta}P_{0}\big\{\widetilde{m}_{H}(X;\theta)\big\}\Big]^{T} \Big[P_{0}\big\{\frac{\partial^{2}}{\partial\theta\partial\theta^{T}}\psi(T,X,Y;\theta)\big\}\Big]^{-1}
\frac{\partial}{\partial\alpha}\psi(T,X,Y;\theta)+\widetilde{m}_{H}(X;\theta)\bigg]\\&&+o_{p}(1),
\end{eqnarray*}
where
$$\widetilde{m}_{B}(X;\theta)=\Phi\Big(\frac{\big(g_{1}(X;\alpha_{1,1})-g_{0}(X;\alpha_{0,1})\big)-c}
{\sqrt{\big(h_{1}(X;\alpha_{1,2})-h_{0}(X;\alpha_{0,2})\big)^{2}+\sigma_{0}^{2}+\sigma_{1}^{2}}}\Big),$$
$$\widetilde{m}_{H}(X;\theta)=\Phi\Big(\frac{\big(g_{0}(X;\alpha_{0,1})-g_{1}(X;\alpha_{1,1})\big)-c}
{\sqrt{\big(h_{0}(X;\alpha_{0,2})-h_{1}(X;\alpha_{1,2})\big)^{2}+\sigma_{0}^{2}+\sigma_{1}^{2}}}\Big),$$
Thus, we can estimate the variances of $\widehat{\mathrm{TBR}}_{c}$ and $\widehat{\mathrm{THR}}_{c}$ by
$$\widehat{\sigma}^{2}_{\mathrm{TBR}_{c}}=\widehat{\mathrm{var}}\bigg[-\Big[P_{n}\big\{\frac{\partial}{\partial \theta}\widetilde{m}_{B}(X;\widehat{\theta})\big\}\Big] ^{T} \Big[P_{n}\big\{\frac{\partial^{2}}{\partial\theta\partial\theta^{T}}\psi(T,X,Y;\widehat{\theta})\big\}\Big]^{-1}
\frac{\partial}{\partial\theta}\psi(T,X,Y;\widehat{\theta})+\widetilde{m}_{B}(X;\widehat{\theta})\bigg]/n,$$
$$\widehat{\sigma}^{2}_{\mathrm{THR}_{c}}=\widehat{\mathrm{var}}\bigg[-\Big[P_{n}\big\{\frac{\partial}{\partial \theta}\widetilde{m}_{H}(X;\widehat{\theta})\big\}\Big] ^{T} \Big[P_{n}\big\{\frac{\partial^{2}}{\partial\theta\partial\theta^{T}}\psi(T,X,Y;\widehat{\theta})\big\}\Big]^{-1}
\frac{\partial}{\partial\theta}\psi(T,X,Y;\widehat{\theta})+\widetilde{m}_{H}(X;\widehat{\theta})\bigg]/n.$$

For the binary case, the asymptotic distribution is
\begin{eqnarray*}
   && \sqrt{n}\big(\widehat{\mathrm{TBR}}-\mathrm{TBR}\big) \\
   &=& \sqrt{n}\bigg[P_{n}-P_{0}\bigg]\bigg[-\Big[\frac{\partial}{\partial \theta}P_{0}\big\{\widetilde{g}_{01}(X;\theta)\big\}\Big]^{T} \Big[P_{0}\big\{\frac{\partial^{2}}{\partial\theta\partial\theta^{T}}\psi(T,X,Y;\theta)\big\}\Big]^{-1}
\frac{\partial}{\partial\theta}\psi(T,X,Y;\theta)+\widetilde{g}_{01}(X;\theta)\bigg]\\&&+o_{p}(1),
\end{eqnarray*}
\begin{eqnarray*}
   && \sqrt{n}\big(\widehat{\mathrm{THR}}-\mathrm{THR}\big) \\
   &=& \sqrt{n}\bigg[P_{n}-P_{0}\bigg]\bigg[-\Big[\frac{\partial}{\partial \theta}P_{0}\big\{\widetilde{g}_{10}(X;\theta)\big\}\Big]^{T} \Big[P_{0}\big\{\frac{\partial^{2}}{\partial\theta\partial\theta^{T}}\psi(T,X,Y;\theta)\big\}\Big]^{-1}
\frac{\partial}{\partial\theta}\psi(T,X,Y;\theta)+\widetilde{g}_{10}(X;\theta)\bigg]\\&&+o_{p}(1),
\end{eqnarray*}
where
$$\widetilde{g}_{01}(X;\theta)=\Phi_{b}\big(\widetilde{\mu}(X;\theta),\widetilde{\Sigma}(X;\theta)\big),$$
$$\widetilde{g}_{10}(X;\theta)=\Phi_{h}\big(\widetilde{\mu}(X;\theta),\widetilde{\Sigma}(X;\theta)\big).$$
Thus, we can estimate the variances of $\widehat{\mathrm{TBR}}$ and $\widehat{\mathrm{THR}}$ by
$$\widehat{\sigma}^{2}_{TBR}=\widehat{\mathrm{var}}\bigg[-\Big[\frac{\partial}{\partial \theta}P_{n}\big\{\widetilde{g}_{01}(X;\theta)\big\}\Big]^{T} \Big[P_{n}\big\{\frac{\partial^{2}}{\partial\theta\partial\theta^{T}}\psi(T,X,Y;\theta)\big\}\Big]^{-1}
\frac{\partial}{\partial\theta}\psi(T,X,Y;\theta)+\widetilde{g}_{01}(X;\theta)\bigg]/n,$$
$$\widehat{\sigma}^{2}_{THR}=\widehat{\mathrm{var}}\bigg[-\Big[\frac{\partial}{\partial \theta}P_{n}\big\{\widetilde{g}_{10}(X;\theta)\big\}\Big]^{T} \Big[P_{n}\big\{\frac{\partial^{2}}{\partial\theta\partial\theta^{T}}\psi(T,X,Y;\theta)\big\}\Big]^{-1}
\frac{\partial}{\partial\theta}\psi(T,X,Y;\theta)+\widetilde{g}_{10}(X;\theta)\bigg]/n.$$